\documentclass[prb,aps,twocolumn,superscriptaddress,showpacs]{revtex4}
\usepackage{amsmath}
\usepackage{amssymb}
\usepackage{amsthm}
\usepackage{amsfonts}
\usepackage{enumerate}
\usepackage{latexsym}
\usepackage{bm}
\usepackage{graphicx}
\usepackage{subfigure}
\usepackage{soul}
\usepackage[dvipsnames]{xcolor}
\usepackage{hyperref}
\hypersetup{colorlinks=true,linkcolor=blue,citecolor=red,urlcolor=RubineRed}
\setstcolor{red}
\usepackage[normalem]{ulem}

\newcommand{\beq}{\begin{equation}}
\newcommand{\eneq}{\end{equation}}

\setstcolor{green}

\input{epsf}

\begin{document}

\title{Floquet engineering of long-range {\it p}-wave superconductivity: Beyond the high-frequency limit}
\author{Zeng-Zhao Li}
\email{zeng-zhao.li@ftf.lth.se}
\affiliation{Division of Solid State Physics and NanoLund, Lund University, Box 118, S-22100 Lund, Sweden}
\affiliation{Max Planck Institute for the Physics of Complex Systems,  N\"othnitzer Strasse 38, 01187 Dresden, Germany}
\affiliation{Quantum Physics and Quantum Information Division, Beijing Computational Science Research Center, Beijing 100094, China}
\author{Chi-Hang Lam}
\email{C.H.Lam@polyu.edu.hk}
\affiliation{Department of Applied Physics, Hong Kong Polytechnic University, Hung Hom, Hong Kong, China}
\author{J. Q. You}
\email{jqyou@csrc.ac.cn}
\affiliation{Quantum Physics and Quantum Information Division, Beijing Computational Science Research Center, Beijing 100094, China}

\begin{abstract}
It has been shown that long-range {\it p}-wave superconductivity in a Kitaev chain can be engineered via an ac field with a  high frequency [Benito \emph{et al.}, Phys. Rev. B {\bf 90}, 205127 (2014)]. For its experimental realization, however, theoretical understanding of Floquet engineering with a broader range of driving frequencies becomes  important. In this paper, focusing on the ac-driven tunneling interactions of a Kitaev chain, we investigate effects from the leading correction to the high-frequency limit on the emergent {\it p}-wave superconductivity. Importantly, we find new engineered long-range {\it p}-wave pairing interactions that can significantly alter the ones in the high-frequency limit at long interaction ranges. We also find that the leading correction additionally generates nearest-neighbor {\it p}-wave pairing interactions with a renormalized pairing energy, long-range tunneling interactions, and in particular multiple pairs of Floquet Majorana edge states that are destroyed in the high-frequency limit.
\end{abstract}

\date{\today}
\maketitle

\section{Introduction}

The Floquet engineering, being a promising quantum engineering and quantum control technology, has attracted much attention in recent years. Its main idea is to design appropriate time-periodic driving protocols to engineer properties of a time-independent effective Hamiltonian that governs the time evolution of periodically driven quantum systems. Besides usual properties analogous to their static counterparts, Floquet quantum systems possess additional unique features that result from the periodicity of their quasienergy spectrum. Given the versatility of driving protocols, Floquet engineering opens up new possibilities for exploring and observing exotic phenomena unreachable in static systems~\cite{Kitagawa10PRB,Rudner13PRX,Jiang11PRL,Kitagawa12NatComm,KunduSeradjeh13PRL,TongAnOh13PRB,Bukov15AdvPhys,EckardtAnisimovas15NJP,Eckardt17RMP,Maricq82PRB,Grozdanov88PRA,Rahav03PRA,CreffieldSols08PRL,Goldman14PRX,Goldman15PRA}. For example, this technique has been employed to realize dynamical localization~\cite{GrossmanHanggi91PRL,Creffield04PRB,GomezPlatero11PRB}, photon-assisted tunneling~\cite{Platero04PhysRep,SiasArimondo08PRL,MaGreiner11PRL}, and novel topological band structures~\cite{Inoue10PRL,Lindner11NatPhys,GomezPlatero13PRL,Delplace13PRB,Grushin14PRL,GomezPlatero114PRB,BenitoPlatero14PRB,Thakurathi13PRB}. 

A promising quantum system for implementing Floquet engineering is a Kitaev chain. It is a one-dimensional {\it p}-wave superconducting wire supporting Majorana zero modes at its ends~\cite{Kitaev01} and can potentially be realized in realistic quantum systems such as a quantum nanowire~\cite{MourikKouwenhoven12Science,Rokhinson12NatPhys,DasShtrikman12NatPhys,DengXu12NanoLett,NadjBernevigYazdani14science,OregRefaelOppen1oPRL,LutchynSauSarma10PRL}.
Similar to the static case, a driven Kitaev chain hosts end states called Floquet Majorana zero modes  
which may realize fault-tolerant quantum computation~\cite{Liu13PRL,ThakurathiKlinovaja17PRB} and also provide a new way to detect these elusive Majorana states~\cite{KunduSeradjeh13PRL,Wang14PRB}.

Interestingly, a very recent work has demonstrated that long-range {\it p}-wave superconductivity (or equivalently long-range many-body spin interactions) can be engineered by periodically driving the tunneling interactions in a Kitaev chain~\cite{BenitoPlatero14PRB}. 
It has been shown for example that the generated next-neighbor interactions become comparable to the first-neighbor interactions as the driving amplitude increases~\cite{BenitoPlatero14PRB}. 
These interesting observations, however, are based on the high driving frequency limit. 
A thorough understanding of Floquet engineering with a broader range of driving frequencies is important for the experimental realization of the effective long-range {\it p}-wave superconductivity. 

In this paper, we consider a driving protocol where the tunneling interactions of a Kitaev chain are modulated by an ac field. We focus on the leading correction to the high-frequency limit. We find that the leading correction results in long-range {\it p}-wave pairing interactions with interaction strengths depending nontrivially on the driving amplitude and frequency of the applied ac field. Compared with the ones corresponding to the high-frequency limit, these new interactions become important when the interaction range increases. We also find that the long-range tunneling interactions, the nearest-neighbor pairing interactions, and multiple Floquet Majorana edge states can now be engineered with the leading correction. We believe that our results are important for the experimental realization of Floquet-engineered long-range {\it p}-wave pairing interactions and tunneling interactions. They also provide new insights about the quantum engineering of new exotic phases in realistic quantum systems.

This paper is organized as follows. In Sec.~\ref{sec:model}, we introduce a model of a time-dependent Kitaev chain. In Sec.~\ref{sec:tunnelling}, we present our driving protocol by considering periodically driven tunneling interactions of a Kitaev chain and derive an effective time-independent Hamiltonian. In Sec.~\ref{sec:LRparing}, we illustrate the important role that the leading correction plays in the engineered long-range pairing interactions. In Sec.~\ref{sec:FloquetMBS}, we demonstrate that multiple Floquet Majorana edge states could be engineered when corrections to the high driving frequency limit are included. Discussions and conclusions are finally given in Sec.~\ref{sec:conclusion}.

\section{The model \label{sec:model}}

The model under our consideration is a Kitaev chain~\cite{Kitaev01} with tunneling interactions which are periodically driven  via, for example, gate voltages applied on a quantum wire. The Hamiltonian of this time-dependent Kitaev chain reads~\cite{BenitoPlatero14PRB} 
\begin{eqnarray}
H\left(t\right)&=&\frac{\mu}{2}\sum_{j=1}^{N} (2f^{\dagger}_j f_j-1) 
-\sum_{j=1}^{N
} \frac{w(t)}{2} (f^{\dagger}_j f_{j+1}+f^{\dagger}_{j+1}f_j ) \notag \\
&& -\sum_{j=1}^{N
} \frac{\Delta}{2} (f^{\dagger}_j f^{\dagger}_{j+1}+f_{j+1}f_j),
\label{eq:realH}
\end{eqnarray}
where $\mu$ and $\Delta$ are chemical potential and pairing energy, respectively, while $w(t)$ is the time-dependent  
tunneling strength. $f_j^{\dagger}$ ($f_j$) is a fermionic operator that creates (annihilates) a fermion on the $j$th site. 
$N$ is the number of the sites. The lattice spacing will be set to unity throughout the paper. This model becomes the well-known Kitaev chain if the Hamiltonian in Eq.~(\ref{eq:realH}) does not depend on time. 

Consider the periodic boundary condition $f_{N+1}=f_1$ and apply the discrete Fourier transformation, $f_k=\frac{1}{\sqrt{N}}\sum_j f_j e^{-ikj}$, 
Eq.~(\ref{eq:realH}) gives the reciprocal-space Hamiltonian
\begin{equation}
H(t)=\sum_{k>0}\Psi^{\dagger}_kH_k(t)\Psi_k,
\end{equation}
where $\Psi^{\dagger}_k=(f^{\dagger}_k, f_{-k})$ is a two-component operator and $H_k(t)$ is the Bogoliubov-de Gennes (BdG) Hamiltonian in the Nambu space given by
\begin{eqnarray}
H_k\left(t\right) 
&=&[\mu-\tilde{w}(t)]\sigma_k^z +\tilde{\Delta} \sigma_k^y.
\label{eq:BdG}
\end{eqnarray}
Here $\sigma_k^{y(z)}$ is the Pauli matrix, $\tilde{w}(t)=w(t)\cos k$, and $\tilde{\Delta}=\Delta \sin k$. For a static Kitaev chain with $w(t)\equiv w_0$, it is known that there is a topological nontrivial phase when $\mu<w_0$ and a trivial phase if $\mu>w_0$ given that $\Delta>0$~\cite{Kitaev01}.
For a dynamical Kitaev chain with the tunneling interactions being periodically driven by an ac field in the limit of a high driving frequency, it has been shown that an interesting effective model with a long-range {\it p}-wave superconductivity (or long-range many-body spin interactions) can be generated~\cite{BenitoPlatero14PRB}.

\section{Driving protocol and effective Hamiltonian of long-range {\it p}-wave superconductivity \label{sec:tunnelling}}

In this section, we present our driving protocol and derive an effective time-independent Hamiltonian in a rotated reference frame~\cite{BenitoPlatero14PRB} which allows the characterization of different topological phases.
Specifically, it is an ac field applied to the tunneling strength in Eq.~(\ref{eq:realH}) given by 
\begin{eqnarray}
w(t) &=& w_{0}+\frac{w_{1}}{2}\cos(\omega t), 
\end{eqnarray}
with $w_{0}$ being a constant, $w_{1}/2$ the driving amplitude, and $\omega$ the driving frequency. Then the BdG Hamiltonian given by Eq.~(\ref{eq:BdG}) becomes 
\begin{eqnarray}
H_k\left(t\right)&=&\{\mu-[w_{0}+\frac{w_{1}}{2}\cos(\omega t)]\cos k\}\sigma_k^z 
+\Delta\sin k\sigma_k^y. \notag \\
\label{eq:BdG_tunnel}
\end{eqnarray}

To obtain a simple effective time-independent Hamiltonian, we need to find a suitable rotating frame. We define
\begin{equation}
S_k^{\dagger}=e^{-i
 \frac{w_{1}}{2\omega}\sin\left(\omega t\right)\cos k\sigma_k^z}, 
\end{equation}
so that Eq.~(\ref{eq:BdG_tunnel}) is transformed as 
\begin{eqnarray}
\tilde{H}_{k}\left(t\right)=S^{\dagger}_{k} (t)H_{k}(t)S_{k}(t)-iS^{\dagger}_{k}(t)\frac{\partial S_{k}(t)}{\partial t},
\end{eqnarray}
which gives
\begin{eqnarray}
\tilde{H}_k(t) &=& [\mu-w_{0}\cos k]\sigma_k^z \notag \\
&& -i\Delta\sin k \sigma_k^{+} e^{-i
\frac{w_{1}}{\omega} \sin(\omega t)\cos k} \notag \\
&& +i \Delta\sin k\sigma_k^{-} e^{i 
\frac{w_{1}}{\omega} \sin(\omega t)\cos k}.  
\end{eqnarray}
By using the Jacobi-Anger expansion 
\begin{eqnarray}
e^{iz\sin\theta}=\sum_{n=-\infty}^{\infty} \mathcal{J}_n(z) e^{in\theta}
\end{eqnarray}
with $\mathcal{J}_n\left(x\right)$ being the $n$th-order Bessel function of the first kind, the Hamiltonian $\tilde{H}_k(t)$ becomes
\begin{eqnarray}
\tilde{H}_k(t)  &=& \sum_{p=-\infty}^{\infty} \tilde{H}_{k,p} e^{ip\omega t},
\label{eq:p}
\end{eqnarray}
where the Fourier components are given by
\begin{eqnarray}
\tilde{H}_{k,p}
&=& (\mu -w_{0}\cos k)\sigma_k^z \delta_{p,0} 
-i\Delta\sin k \sigma_k^{+} \mathcal{J}_{-p} \notag \\
&& +i\Delta\sin k\sigma_k^{-} \mathcal{J}_{p},
\label{eq:FourierComp-tunn}
\end{eqnarray}
with
\begin{equation}
\mathcal{J}_{\pm p} \equiv \mathcal{J}_{\pm p}  \left(\frac{w_{1}}{\omega} \cos k \right). \label{eq:Jp}
\end{equation}

Then one can define the one-period time-evolution operator $U\left(T,0\right)=e^{-i\tilde{H}_k^{\mathrm{eff}}T}$ with the period $T=2\pi/\omega$ and a time-independent effective Hamiltonian $\tilde{H}_k^{\mathrm{eff}}$. Using the Magnus expansion~\cite{Blanes09PhysRep} and Eq. (\ref{eq:FourierComp-tunn}), the effective Hamiltonian is expressed as a power series expansion in $1/\omega$~\cite{Rahav03PRA,Bukov15AdvPhys,EckardtAnisimovas15NJP,Eckardt17RMP}, namely
\begin{eqnarray}
\tilde{H}_k^{\mathrm{eff}}&=&\tilde{H}_{k,0}+\frac{1}{\omega}[\tilde{H}_{k,0},\tilde{H}_{k,1}]-\frac{1}{\omega}[\tilde{H}_{k,0},\tilde{H}_{k,-1}] \notag \\
&& -\frac{1}{\omega}[\tilde{H}_{k,-1},\tilde{H}_{k,1}]+\cdots.
\label{eq:effectiveH}
\end{eqnarray}
The convergence condition for the expansion is $\int_0^T \parallel \tilde{H}_k(t)\parallel dt < \pi$ where $\parallel\cdots\parallel$ denotes the Euclidean norm~\cite{BenitoPlatero14PRB,Blanes09PhysRep,Feldman84PLA}. 

The previously considered high-frequency limit corresponds to  $\omega\rightarrow\infty$ in Eq.~({\ref{eq:effectiveH}}) or equivalently taking only the $p=0$ term in Eq.~(\ref{eq:p}). In contrast, we also include the leading correction by considering all four terms on the right-hand side of Eq.~({\ref{eq:effectiveH}}) and obtain an effective BdG Hamiltonian (see Appendix~\ref{sec:AppendA})
\begin{eqnarray}
\tilde{H}_k^{\mathrm{eff}} 
&=& \left(\mu- w_{0}\cos k +\frac{4\mathcal{J}_0\mathcal{J}_{1}}{\omega}  \Delta^2 \sin^2 k \right)\sigma_k^z \notag \\
&&+ \left[ (\mathcal{J}_{0} -\frac{4\mu}{\omega} )\Delta \sin k  
+\frac{2w_{0} \mathcal{J}_{1} }{\omega} \Delta \sin(2k)\right] \sigma_k^{y}. 
\notag \\
\label{eq:bdg_tunnel_Bessel}
\end{eqnarray}
Here $\mathcal{J}_0$ and $\mathcal{J}_1$ are given by Eq.~(\ref{eq:Jp}). This effective Hamiltonian in Eq.~(\ref{eq:bdg_tunnel_Bessel}) reduces to Eq.~(31) of Ref.~\onlinecite{BenitoPlatero14PRB} in the high-frequency limit. 
It contains in particular a pairing term which reads, after summing over the Fourier modes [see also Eq.~({\ref{eq:paritingV}}) in Appendix~\ref{sec:AppendA}],
\begin{eqnarray}
V&=&-i\sum_{k>0} \Big\{ (\mathcal{J}_{0} -\frac{4\mu}{\omega} )\Delta \sin k \notag \\
&&
+\frac{2w_{0} \mathcal{J}_{1} }{\omega} \Delta \sin(2k)   \Big\} 
(f_{k}^{\dagger} f_{-k}^{\dagger} -f_{-k}f_{k} ).
\end{eqnarray}

To have a better understanding of this effective Hamiltonian, it will be converted to the real space by performing the inverse Fourier transformation. This is however non-trivial due to the $k$-dependent argument of the Bessel functions. We thus first evaluate the Bessel functions by using the expansions~\cite{Olver10,Watson44}
\begin{eqnarray}
\mathcal{J}_0 &=& \sum_{m=0}^{\infty} \frac{(-1)^m}{(m!)^2}  \left(\frac{w_{1} \cos k}{2\omega}\right)^{2m}, \label{eq:Bessel0} \\
\mathcal{J}_{1} 
&=& \sum_{m=0}^{\infty} \frac{(-1)^m}{m! (m+ 1)!} \left(\frac{w_{1} \cos k}{2\omega}\right)^{2m+ 1}, \label{eq:Bessel1} \\
\mathcal{J}_0\mathcal{J}_{1} &=& \sum_{m=0}^{\infty} \frac{\left(-1\right)^m \left(2m+1\right)}{[m!(m+1)!]^2} \left(\frac{w_{1} \cos k}{2\omega}\right)^{2m+1}.\label{eq:BesselProd}
\end{eqnarray}
After substituting into Eq.~(\ref{eq:bdg_tunnel_Bessel}) and performing some simplifications, the effective BdG Hamiltonian becomes
\begin{eqnarray}
\tilde{H}_k^{\mathrm{eff}} 
&=& \left[\mu- w_{0}\cos k +\sum_{m=0}^{\infty} \sum_{r=1,3,\cdots}^{2m+3} 2\mathcal{C}_1 \mathcal{D}_1 \cos(kr) \right]\sigma_k^z \notag \\
&& +\left[ \sum_{m=0}^{\infty} \sum_{r^{\prime}=1,3,\cdots}^{2m+1} 2 \mathcal{C}_2 \mathcal{D}_2 \sin(kr^{\prime}) -\frac{4\mu\Delta}{\omega} \sin k  \right. \notag\\
&& \left. +\sum_{m=0}^{\infty} \sum_{r=1,3,\cdots}^{2m+3} 2\mathcal{C}_3 \mathcal{D}_3 \sin(kr) \right] \sigma_k^{y}, 
\label{eq:bdg_tunnel_Bessel_simplified}
\end{eqnarray}
where
\begin{eqnarray}
\mathcal{C}_1 &=& 2 C_{2m+1}^{m-\frac{r-1}{2}} - C_{2m+1}^{m-\frac{r-3}{2}} -C_{2m+1}^{m-\frac{r+1}{2}}, \label{eq:C1} \\
\mathcal{C}_2 &=& C_{2m}^{m-\frac{r^{\prime}-1}{2}} -C_{2m}^{m-\frac{r^{\prime}+1}{2}}, \label{eq:C2} \\
\mathcal{C}_3 &=& C_{2m+1}^{m-\frac{r-3}{2}} -C_{2m+1}^{m-\frac{r+1}{2}}, \label{eq:C3}
\end{eqnarray}
and
\begin{eqnarray}
\mathcal{D}_1 &=& \frac{\Delta^2}{\omega} \frac{(-1)^m (2m+1)}{[m!(m+1)!]^2} \left(\frac{w_1}{4\omega}\right)^{2m+1}, \label{eq:D1} \\
\mathcal{D}_2 &=& \frac{(-1)^m \Delta}{2(m!)^2} \left(\frac{w_1}{4\omega}\right)^{2m}, \label{eq:D2} \\
\mathcal{D}_3 &=& \frac{w_0\Delta}{\omega} \frac{(-1)^m}{m!(m+1)!} \left(\frac{w_1}{4\omega}\right)^{2m+1}. \label{eq:D3}
\end{eqnarray}
Here, $C_{a}^{b}$ with $b>0$ is a binomial coefficient and $C_{2m+1}^{-1}=C_{2m+1}^{-2}=C_{2m}^{-1}=0$. Besides the original $\cos k$-dependent term, the coefficient of $\sigma_k^z$ in Eq.~(\ref{eq:bdg_tunnel_Bessel_simplified}) now attains other terms involving $\cos(kr)$. It will be evident from the real-space Hamiltonian to be derived below that those with $r>1$ imply long-range tunnelings. In addition, the $\sigma_k^y$ term in Eq.~(\ref{eq:bdg_tunnel_Bessel_simplified}) clearly demonstrates  effective superconducting pairing interactions with coefficients $4\mu\Delta \sin k /\omega$, $2 \mathcal{C}_2 \mathcal{D}_2 \sin(kr^{\prime})$, and $2\mathcal{C}_3 \mathcal{D}_3 \sin(kr)$. Being independent of $1/\omega$, the term involving $2 \mathcal{C}_2 \mathcal{D}_2 \sin(kr^{\prime})$ was previously obtained in the high-frequency limit (i.e., $\omega\rightarrow\infty$)~\cite{BenitoPlatero14PRB}. However, those involving $4\mu\Delta\sin k/\omega$ and $2\mathcal{C}_3 \mathcal{D}_3 \sin(kr)$ are proportional to $1/\omega$ and represent the leading corrections to the high frequency limit. In particular, terms involving $2\mathcal{C}_3 \mathcal{D}_3 \sin(kr)$ provide a correction to those of $2 \mathcal{C}_2 \mathcal{D}_2 \sin(kr^{\prime})$. 

Here are several remarks. From Eqs.~(\ref{eq:D1}) and (\ref{eq:D3}), it is clear that the reported $\mathcal{D}_1$ and $\mathcal{D}_3$ depend on the amplitude $w_{1}$, the frequency $\omega$ of the driving field and the pairing energy $\Delta$. These nontrivial dependences on the applied electric field may provide flexible controllability. For $\omega\rightarrow\infty$, $\mathcal{D}_1$, $\mathcal{D}_2$ and $\mathcal{D}_3$ in general vanish, with a notable exception that $\mathcal{D}_2=\Delta/2$ at $m=0$.

\begin{figure}
\centering
\begin{minipage}[c]{0.48\textwidth}
  \centering
  \includegraphics[width=0.9\columnwidth]{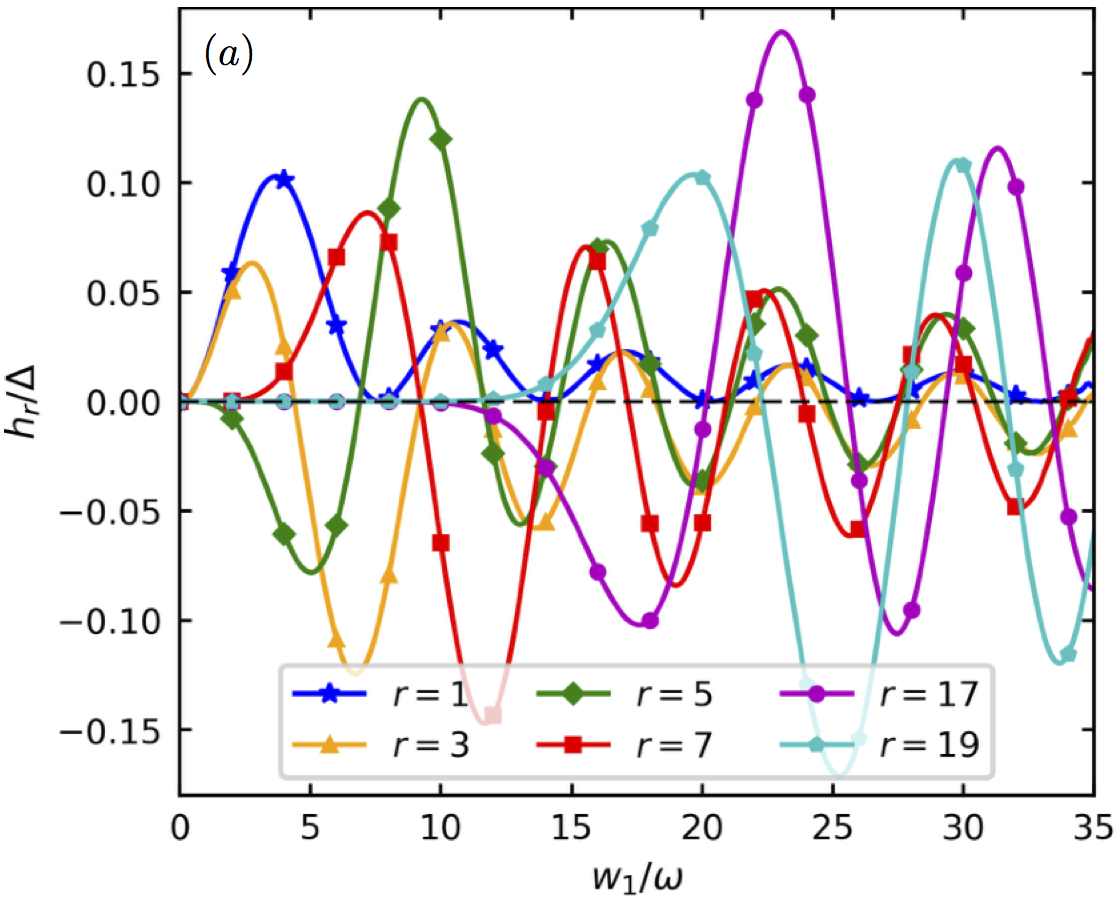}
\end{minipage} \newline
\begin{minipage}[c]{0.49\textwidth}
  \centering
  \includegraphics[width=0.9\columnwidth]{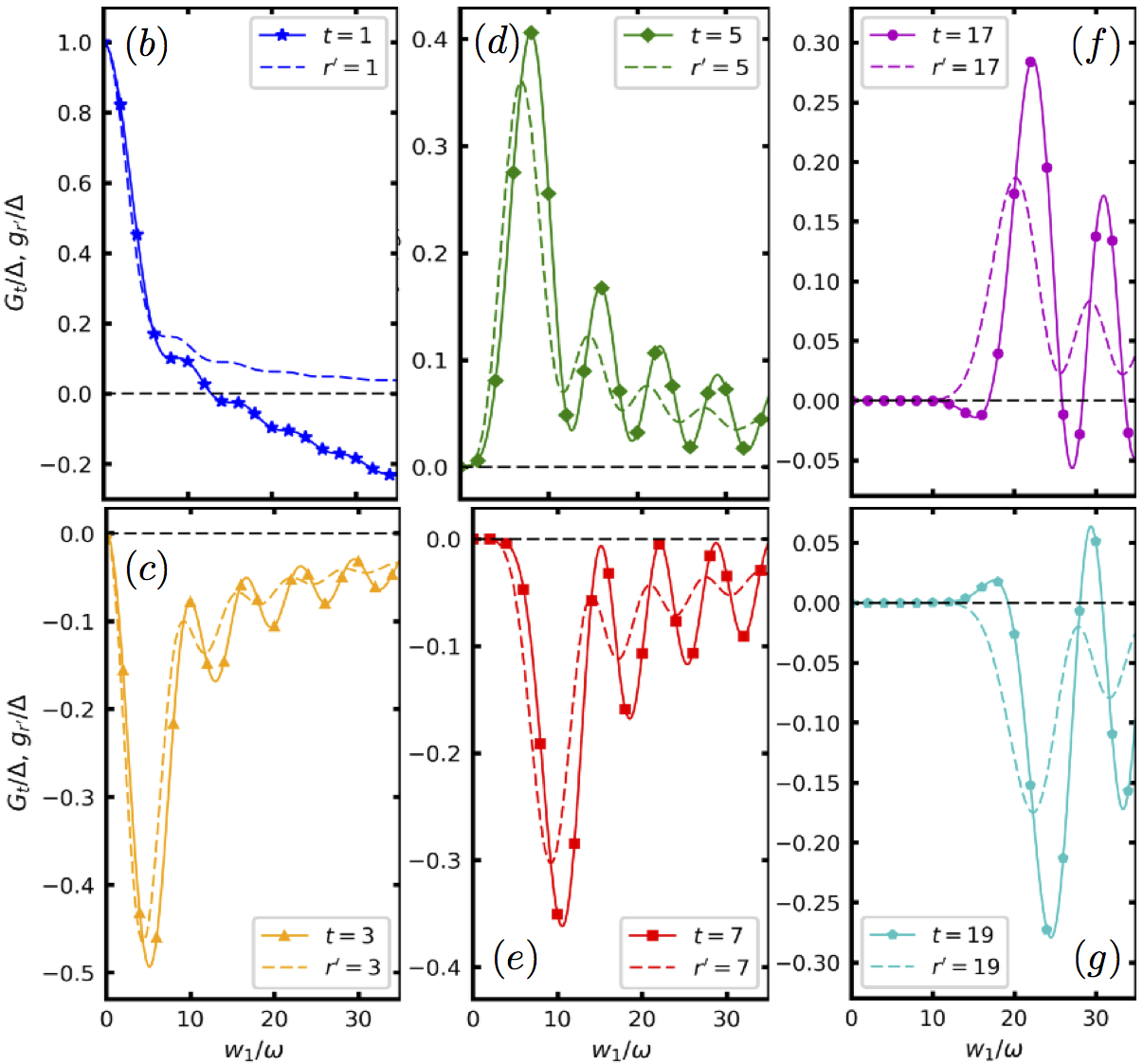}
\end{minipage}
\caption{The superconducting pairing strength (a) $h_r/\Delta$ and (b)-(g) $g_{r^{\prime}}/\Delta$ (dashed curves) without correction and $G_t/\Delta$ (solid curves) with correction for different interaction ranges $r, t, r^{\prime}\in\{1,3,5,7,17,19\}$. The parameters are $w_0=0.038w_1$ and $\mu=0.002w_1$.}
\label{fig:BCSstrength}
\end{figure}

We now perform an inverse Fourier transformation to  ${H}^{\mathrm{eff}}=\sum_{k>0}\Psi^{\dagger}_k \tilde{H}_k^{\mathrm{eff}}(t)\Psi_k$ with $\tilde{H}_k^{\mathrm{eff}}$ given by Eq.~(\ref{eq:bdg_tunnel_Bessel_simplified}) and obtain 
\begin{eqnarray}
\tilde{H}^{\mathrm{eff}} &=& \sum_j \Big\{\frac{\mu}{2}(2 f_j^{\dagger}f_j-1)
- \frac{w_{0}}{2}(f_j^{\dagger}f_{j+1} + f_{j+1}^{\dagger}f_j) \notag \\
&&
+\frac{2\mu\Delta}{\omega} (f_j^{\dagger} f_{j+1}^{\dagger} +f_{j+1} f_j) \notag \\
&& + \sum_{m=0}^{\infty} \sum_{r=1,3,\cdots}^{2m+3} \mathcal{C}_1 \mathcal{D}_1 
 (f_j^{\dagger}f_{j+r} +f_{j+r}^{\dagger}f_{j})  \notag \\
&& 
-\sum_{m=0}^{\infty} \sum_{r^{\prime}=1,3,\cdots}^{2m+1} 
\mathcal{C}_2 \mathcal{D}_2 (f_j^{\dagger}f_{j+r^{\prime}}^{\dagger} +f_{j+r^{\prime}} f_{j}) \notag \\
&& - \sum_{m=0}^{\infty} \sum_{r=1,3,\cdots}^{2m+3} \mathcal{C}_3 \mathcal{D}_3 
(f_{j}^{\dagger} f_{j+r}^{\dagger} +f_{j+r} f_{j})
\Big\}. 
\label{eq:rs_H_eff}
\end{eqnarray}
The first two terms in the curly brackets representing the on-site chemical potential and nearest-neighbor tunnelings,  
are the same as those in the static counterpart of Eq.~(\ref{eq:realH}). The fifth term involving $\mathcal{C}_2\mathcal{D}_2$ includes long-range {\it p}-wave pairing interactions, e.g., $f_j^{\dagger} f^{\dagger}_{j+r^{\prime}}$ ($r^{\prime}=3, 5, \cdots, 2m+1$) and is also present in the high-frequency limit~\cite{BenitoPlatero14PRB}. 

We now focus on new correction terms of leading order $1/\omega$. One is the third term in the curly brackets in Eq.~(\ref{eq:rs_H_eff}). It represents nearest-neighbor superconducting pairing with a renormalized pairing energy $2\mu\Delta/\omega$ that depends on the driving frequency $\omega$ in contrast to a constant bare value of $\Delta/2$ in Eq.~(\ref{eq:realH}). 
The other is the fourth term representing both engineered nearest-neighbor ($r=1$) and long-range ($r\ge3$) tunneling $f_j^{\dagger}f_{j+r}$ with a  strength $\mathcal{C}_1\mathcal{D}_1/4$.  
More interestingly, the sixth term involving $\mathcal{C}_3\mathcal{D}_3f_{j}^{\dagger} f_{j+r}^{\dagger}/2$ ($r=1,3,\cdots,2m+3)$ is the leading correction to the {\it p}-wave pairing. These new long-range pairing corrections can  both enhance or reduce the pairing interaction strength as demonstrated below. 
Finally, we emphasize that these corrections are all proportional to $1/\omega$ or its powers and vanish in the high frequency limit.

\begin{figure*}%
\centering
  \includegraphics[width=1.7\columnwidth]{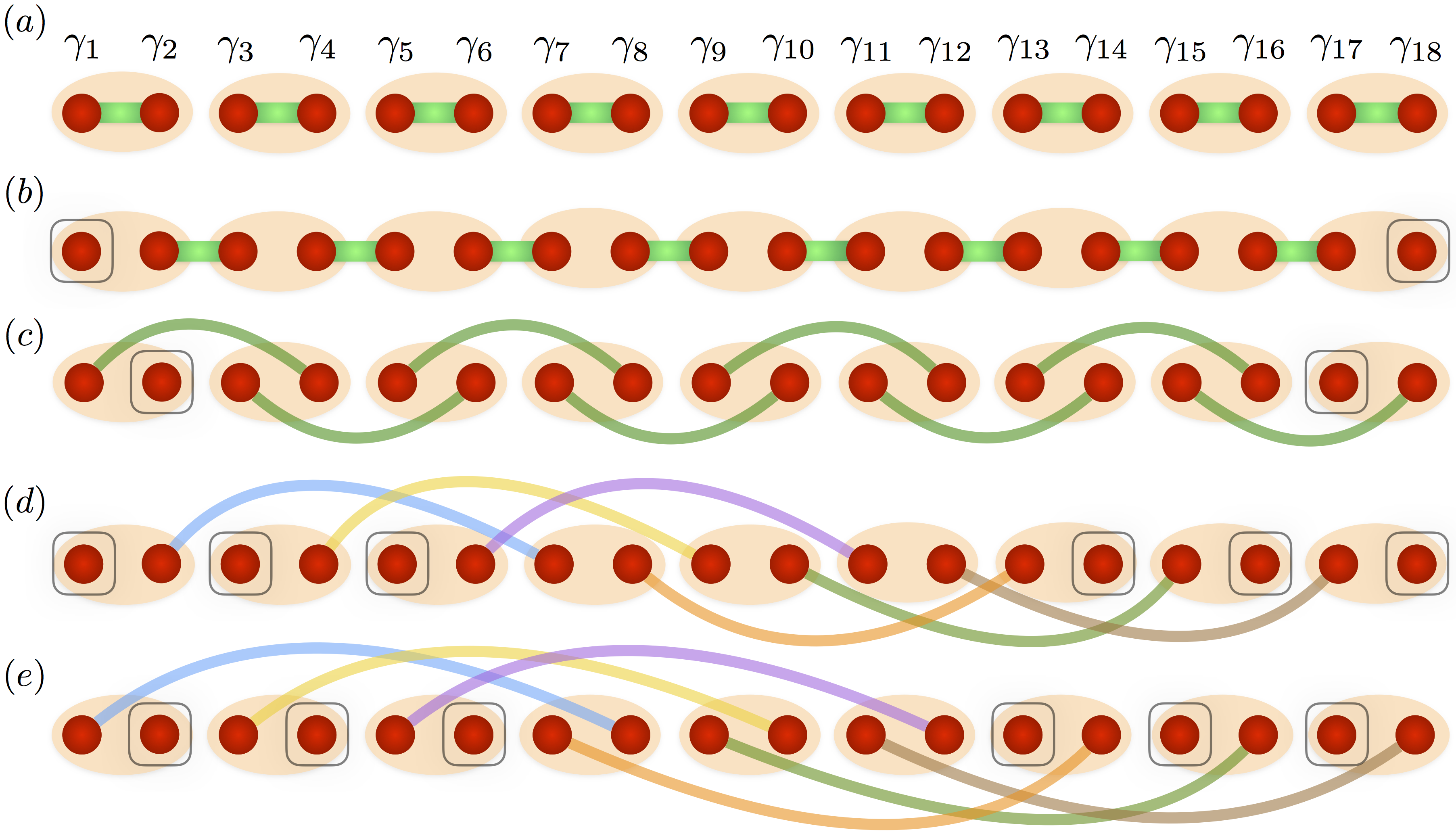} 
\caption{Schematic diagram of Floquet Majorana edge states. The dominant interaction term and the resulting edge states are (a) $i\gamma_{2j-1}\gamma_{2j}$ and no edge states, (b) $i\gamma_{2j}\gamma_{2j+1}$ and ($\gamma_1$, $\gamma_{2N}$), (c) $i\gamma_{2j-1}\gamma_{2j+2}$ and ($\gamma_2$, $\gamma_{2N-1}$), (d) $i \gamma_{2j}\gamma_{2j+2r-1}$ and $(\gamma_{2l-1}, \gamma_{2N-2(l-1)})$, and (e) $i \gamma_{2j-1}\gamma_{2j+2r}$ and $(\gamma_{2l}, \gamma_{2N-(2l-1)})$ with $1\le l\le r$ and $r=1,3,\cdots$. In order to illustrate the long-range {\it p}-wave pairing ($r>1$), here we consider $r=3$ and at least $2N=18$ sites. Note that (a), (b), and (c) also appear for a Kitaev chain with appropriate parameters even without applying an ac field. The box is used to distinguish a decoupled zero-energy Majorana.}
\label{fig:schematic}
\end{figure*}

\section{Long-range {\it p}-wave pairing \label{sec:LRparing}}

To illustrate that corrections to the high-frequency limit can play an important role in the engineered long-range {\it p}-wave superconductivity, we now analyze in detail the pairing strength. Given an interaction range $t$, the pairing term in Eq.~(\ref{eq:bdg_tunnel_Bessel_simplified}) can be written as $V=\sum_{t} G_t \sin(kt) \sigma_k^y$ with 
\begin{eqnarray}
G_t &=& f_1 \delta_{1,t} +g_{r\prime}\delta_{r\prime,t}  +h_r \delta_{r,t}, \label{eq:G}
\end{eqnarray}
where
\begin{eqnarray}
f_1 (\mu)&=& -\frac{4\mu\Delta}{\omega}, \label{eq:f} \\
g_{r\prime} (w_1)&=& \sum_{m=0}^{\infty} 2\mathcal{C}_2\mathcal{D}_2, \label{eq:g}\\
h_{r} (w_1)&=& \sum_{m=0}^{\infty} 2\mathcal{C}_3\mathcal{D}_3 \label{eq:h}.
\end{eqnarray}
Here $t,r^{\prime},r\in\{1,3,\cdots\}$. Note that $g_{r\prime} (w_1)$ is the only non-vanishing term in the high frequency limit and is identical to the previous result in Ref.~\onlinecite{BenitoPlatero14PRB}. In contrast, $f_1$ and $h_r$ in Eqs.~(\ref{eq:f}) and (\ref{eq:h}) respectively are corrections reported for the first time in this paper. In particular, $h_r (r>1)$ denotes the long-range pairing correction, at which we are particularly interested.

Figure~\ref{fig:BCSstrength}(a) shows that the correction $h_r$ is negligible for small values of $w_1/\omega\ll1$. However, with the increase of $w_1/\omega$, the individual terms $\lvert h_r\rvert$ increase and further show damping oscillations. The impacts due to the corrections $h_r$ and $f_{1}$ are further illustrated in Figs.~\ref{fig:BCSstrength}(b)-(g) that show
the superconducting pairing strengths $G_t$ and $g_{r\prime}$ with and without corrections, respectively.
For the nearest-neighbor pairing interaction with $t=r^{\prime}=1$ in Fig.~\ref{fig:BCSstrength}(b), $G_1$ shows a damping washboard behavior and changes from positive to negative in addition to exhibiting damping oscillations, in contrast to $g_{1} $ which approaches to zero.
Importantly, for the long-range pairing, e.g., $t,r^{\prime}\ge3$ in Figs.~\ref{fig:BCSstrength}(c)-(g), it is shown that the deviation of $g_{r\prime}$ from $G_t$ 
becomes significant when increasing the interaction length, indicating the importance of the correction at long interaction ranges. For example, we clearly observe shifted extremum points and also enhanced magnitudes as the range increases from $7$ to $19$ in Figs.~\ref{fig:BCSstrength}(e) and (g), respectively. 
It is also shown that long-range pairing interactions $G_t (t>1)$ attain magnitudes comparable to or beyond that of $G_1$ for $w_1/\omega\agt 1$.
Interestingly, we also find that there are regimes where one long-range pairing strength can be larger than that of the other ranges, e.g., $\vert G_7\vert>\vert G_{t}\vert$ ($t\neq7$), implying the possibility of the engineering of a predominantly arbitrarily-ranged pairing interaction. In a nutshell, the long-range pairing corrections are crucial for any detectable long-range {\it p}-wave superconductivity.

\section{Floquet Majorana edge states \label{sec:FloquetMBS}}

Although the long-range {\it p}-wave pairing was reported previously in the limit of a high driving frequency (i.e., $\omega\rightarrow\infty$), edge states and their observability were not discussed~\cite{BenitoPlatero14PRB}. 
In this section, we are going to show that the leading correction of order 
$1/\omega$ gives rise to multiple pairs of Floquet Majorana edge states which, however, are destroyed in the high-frequency limit.

Considering open boundary conditions and applying Majorana operators
\begin{equation}
\gamma_{2j-1}=f_j+f_j^{\dagger}, ~~~ \gamma_{2j}=-i(f_j-f_j^{\dagger}),
\label{eq:MFopera}
\end{equation}
Eq.~(\ref{eq:rs_H_eff}) is represented as
\begin{eqnarray}
\tilde{H}^{\mathrm{eff}} &=& \sum_j \Big\{\frac{\mu}{2} i \gamma_{2j-1}\gamma_{2j} 
+\frac{w_{0}\omega-4\mu\Delta}{4\omega} i \gamma_{2j}\gamma_{2j+1} \notag\\
&& -\frac{w_{0}\omega +4\mu\Delta}{4\omega} i \gamma_{2j-1} \gamma_{2j+2}\notag \\
&& -  \sum_{m=0}^{\infty} \Big[ \sum_{r=1,3,\cdots}^{2m+3} 
\Big(\frac{\mathcal{C}_1 \mathcal{D}_1 -\mathcal{C}_3 \mathcal{D}_3}{2} 
i \gamma_{2j}\gamma_{2j+2r-1} \notag \\
&& -\frac{\mathcal{C}_1 \mathcal{D}_1 +\mathcal{C}_3 \mathcal{D}_3}{2} 
i \gamma_{2j-1}\gamma_{2j+2r} \Big)  \notag \\
&& 
- \sum_{r^{\prime}=1,3,\cdots}^{2m+1} 
\frac{\mathcal{C}_2 \mathcal{D}_2}{2} i\Big(
\gamma_{2j}\gamma_{2j+2r^{\prime}-1}
+ \gamma_{2j-1}\gamma_{2j+2r^{\prime}} \Big) \Big] \Big\}. \notag\\
\label{eq:MajoranaH}
\end{eqnarray}

\subsection{High-frequency limit}

\begin{figure}
\centering
  \includegraphics[width=.93\columnwidth]{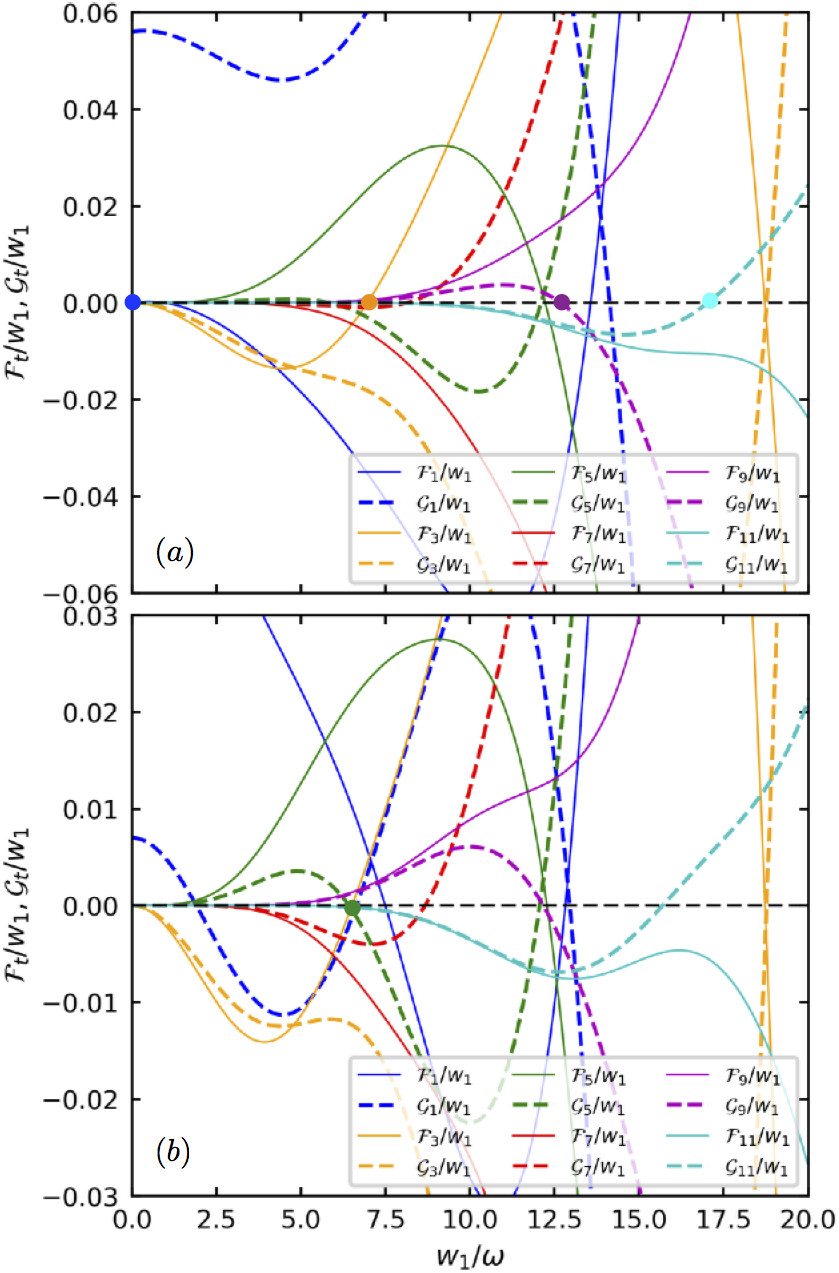} 
\caption{The coupling strengths $\mathcal{F}_t$ and $\mathcal{G}_t$ for interactions $i \gamma_{2j-1}\gamma_{2j+2t}$ and $i \gamma_{2j}\gamma_{2j+2t-1}$ respectively as a function of $w_1/\omega$. The blue dot in (a) indicates a single Floquet Majorana pair while other dots represent the solutions to the equations $\mathcal{F}_t\mathcal{G}_t=0$ and $\mathcal{F}_t+\mathcal{G}_t\neq0$ necessary for the existence of multiple Majorana pairs. The green dot in (b) denotes the existence of multiple pairs of Floquet Majorana states. We take $w_0=0.028w_1$, $\mu=-0.01w_1$ in (a) and $w_0=-0.021w_1$, $\mu=-0.00001w_1$ in (b). Other parameter is $\Delta=0.028w_1$.}
\label{fig:MBSsolution}
\end{figure}

In order to better understand the important role that the leading correction plays in the generation of detectable Majorana edge states, let us first start with 
the high-frequency limit. The above Hamiltonian in Eq.~(\ref{eq:MajoranaH}) is further simplified to
\begin{eqnarray}
\tilde{\tilde{H}}^{\mathrm{eff}} &=& \sum_j \Big\{\frac{\mu}{2} i \gamma_{2j-1}\gamma_{2j} 
+\frac{w_{0}}{4} i \Big(\gamma_{2j}\gamma_{2j+1}  
- \gamma_{2j-1} \gamma_{2j+2} \Big) \notag \\
&& 
+\sum_{r^{\prime}=1,3,\cdots}^{2m+1} 
\frac{\mathcal{C}_2 \mathcal{D}_2}{2} i\Big(
\gamma_{2j}\gamma_{2j+2r^{\prime}-1}
+\gamma_{2j-1}\gamma_{2j+2r^{\prime}} \Big) \Big] \Big\}. \notag\\
\label{eq:MajoranaH_highfreq}
\end{eqnarray}
In the second term on the right-hand side of Eq.~(\ref{eq:MajoranaH_highfreq}), two Majoranas  $\gamma_1$ and $\gamma_{2N}$ that are absent from the interaction $i\gamma_{2j}\gamma_{2j+1}$ [see  Fig.~\ref{fig:schematic}(b)] are however coupled via the interaction $i\gamma_{2j-1}\gamma_{2j+2}$ to $\gamma_4$ and $\gamma_{2N-3}$ [see Fig.~\ref{fig:schematic}(c)], respectively. These couplings destroy the Majoranas. Also, $\gamma_{2}$ and $\gamma_{2N-1}$ absent from $i\gamma_{2j-1}\gamma_{2j+2}$ [see Fig.~\ref{fig:schematic}(c)] are  
coupled via the interaction $i\gamma_{2j}\gamma_{2j+1}$ to $\gamma_3$ and $\gamma_{2N-2}$ [see Fig.~\ref{fig:schematic}(b)], respectively. 
Due to the third term in Eq.~(\ref{eq:MajoranaH_highfreq}), Majorana pairs $(\gamma_{2l-1}, \gamma_{2N-2(l-1)})$ with $1\le l\le r^{\prime}$,
i.e., $\gamma_1, \gamma_3, \cdots, \gamma_{2r^{\prime}-1}$ on one end of the chain and $\gamma_{2N-2(r^{\prime}-1)}, \cdots, \gamma_{2N-2}, \gamma_{2N}$ on the other end [see Fig.~\ref{fig:schematic}(d) where $r^{\prime}=3$ and $2N=18$],  
which are not coupled by $i \gamma_{2j}\gamma_{2j+2r^{\prime}-1}$, are on the other hand coupled by $i \gamma_{2j-1}\gamma_{2j+2r^{\prime}}$. 
Similarly, for the interaction $i \gamma_{2j-1}\gamma_{2j+2r^{\prime}}$, we have  
Majorana pairs $(\gamma_{2l}, \gamma_{2N-(2l-1)})$, namely, $\gamma_2, \gamma_4, \cdots, \gamma_{2r^{\prime}}$ on one end and $\gamma_{2N-(2r^{\prime}-1)}, \cdots, \gamma_{2N-3}, \gamma_{2N-1}$ on the other end as shown in Fig.~\ref{fig:schematic}(e), which, however, are coupled by the interaction $i \gamma_{2j}\gamma_{2j+2r^{\prime}-1}$.
The unavoidable destruction of Majorana edge states according to the above analysis essentially results from the same strengths for two kinds of interactions. When the second and third terms are comparable in magnitudes, it is possible to have two unequal coupling strengths $\mathcal{C}_2\mathcal{D}_2/2+ w_0/4$ and $\mathcal{C}_2\mathcal{D}_2/2- w_0/4$ with $r^{\prime}=1$ for interactions $i\gamma_{2j}\gamma_{2j+1}$ and $i\gamma_{2j-1} \gamma_{2j+2}$, respectively, implying that there would be a single Majorana pair $(\gamma_1,\gamma_{2N})$ or $(\gamma_{2}, \gamma_{2N-1})$ as indicated by the blue dot in Fig.~\ref{fig:MBSsolution}(a) [discussions about Fig.~\ref{fig:MBSsolution}(a) are given below]. 
As a result, in the high-frequency limit with equal long-range interaction strengths (e.g., $\mathcal{C}_2\mathcal{D}_2/2$ for $r^{\prime}>1$), there cannot exist multiple pairs of Floquet Majorana edge states.

\begin{figure}
\centering
  \includegraphics[width=.93\columnwidth]{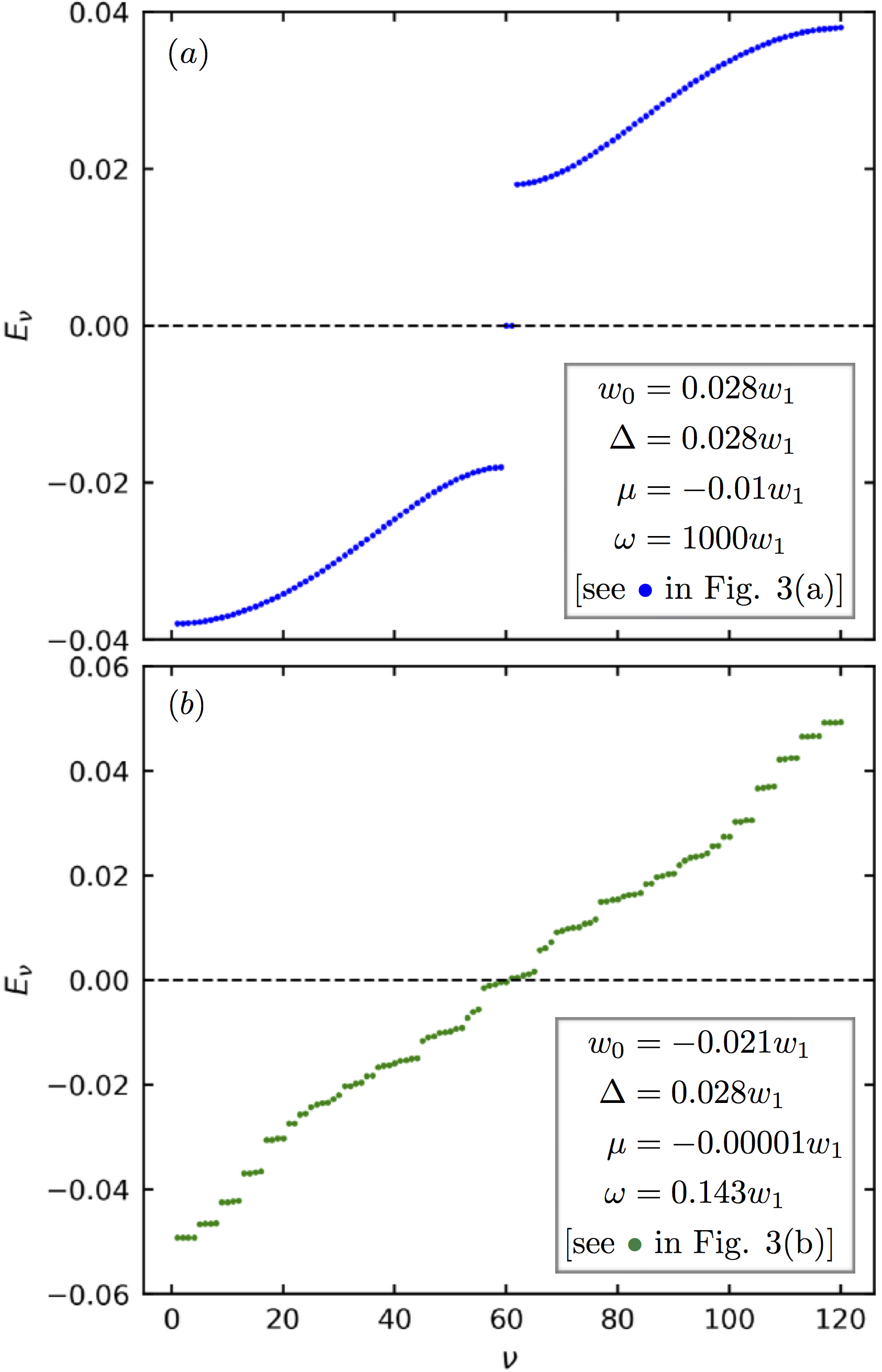} 
\caption{The energy spectrum $E_{\nu}$ of the effective Hamiltonian [i.e., Eq.~(\ref{eq:rs_H_eff})] with $N=60$ sites versus the index $\nu$. We take $w_1/\omega=0.001$ in (a) and $6.989$ in (b) corresponding to the blue dot in Fig.~\ref{fig:MBSsolution}(a) and the green dot in Fig.~\ref{fig:MBSsolution}(b) respectively. Other parameters are the same as those in Fig.~\ref{fig:MBSsolution}.}
\label{fig:MBSspectrum}
\end{figure}

\subsection{Beyond the high driving frequency}

We now go back to consider Eq.~(\ref{eq:MajoranaH}) that includes the leading correction and therefore goes beyond the high driving frequency limit. Let us first consider the second and third terms inside the curly brackets in Eq.~(\ref{eq:MajoranaH}). Performing similar analysis to that on Eq.~(\ref{eq:MajoranaH_highfreq}), we know that it is possible to have two Majorana edge states $\gamma_1$ and $\gamma_{2N}$ if $w_0\omega \approx -4\mu\Delta$ or $\gamma_{2}$ and $\gamma_{2N-1}$ if $w_0\omega \approx 4\mu\Delta$, as shown in Fig.~\ref{fig:schematic}(b) or (c) respectively. This simple condition for having Majoranas is based on the assumption of negligible contributions from the fourth, fifth and sixth terms with $r=r^{\prime}=1$. 
When the fourth and fifth terms including long-range interactions (e.g., $r\ge3$) become dominant, there would be multiple Majorana pairs $(\gamma_{2l-1}, \gamma_{2N-2(l-1)})$ if $\mathcal{C}_1\mathcal{D}_1\approx -\mathcal{C}_3\mathcal{D}_3$ and  $(\gamma_{2l}, \gamma_{2N-(2l-1)})$ if $\mathcal{C}_1\mathcal{D}_1\approx \mathcal{C}_3\mathcal{D}_3$ ($1\le l\le r$ with $r=1,3,\cdots,2m+3$).

Similar to the analyses above, 
we can extract from Eq.~(\ref{eq:MajoranaH}) a general condition for the existence of Floquet Majorana edge states. The coupling strengths for interactions $i \gamma_{2j-1}\gamma_{2j+2t}$ and $i \gamma_{2j}\gamma_{2j+2t-1}$ are defined by $\mathcal{F}_t$ and $\mathcal{G}_t$ respectively, which are given by
\begin{eqnarray}
\mathcal{F}_t &=& G_t +F_t, \label{eq:mathcalF}\\
\mathcal{G}_t &=& G_t -F_t, \label{eq:mathcalG}
\end{eqnarray}
where the pairing strength $G_t$ is given by Eq.~(\ref{eq:G}) and the tunneling strength $F_t$ is defined as 
\begin{eqnarray}
F_t &=& p_1 \delta_{1,t}+q_r\delta_{r,t}, 
\end{eqnarray}
with
\begin{eqnarray}
p_1 &=& -w_0, \\
q_r(w_1) &=& \sum_{m=0}^{\infty} 2\mathcal{C}_1\mathcal{D}_1.
\end{eqnarray}
Therefore, Majorana edge states are indicated by the solutions to equations $\mathcal{F}_t\mathcal{G}_t=0$ and $\mathcal{F}_t+\mathcal{G}_t\neq0$. 
If $\mathcal{F}_t=0$ and $\mu\ll\vert\mathcal{G}_t\rvert\neq0$, we have Majorana edge states $\gamma_{2l-1}$ and $\gamma_{2N-2(l-1)}$. Alternatively, if $\mu\ll\lvert\mathcal{F}_t\rvert\neq0$ and $\mathcal{G}_t=0$, the edge states become $\gamma_{2l}$ and $\gamma_{2N-(2l-1)}$ with $1\le l\le t$.  

\begin{figure}
\centering
  \includegraphics[width=.94\columnwidth]{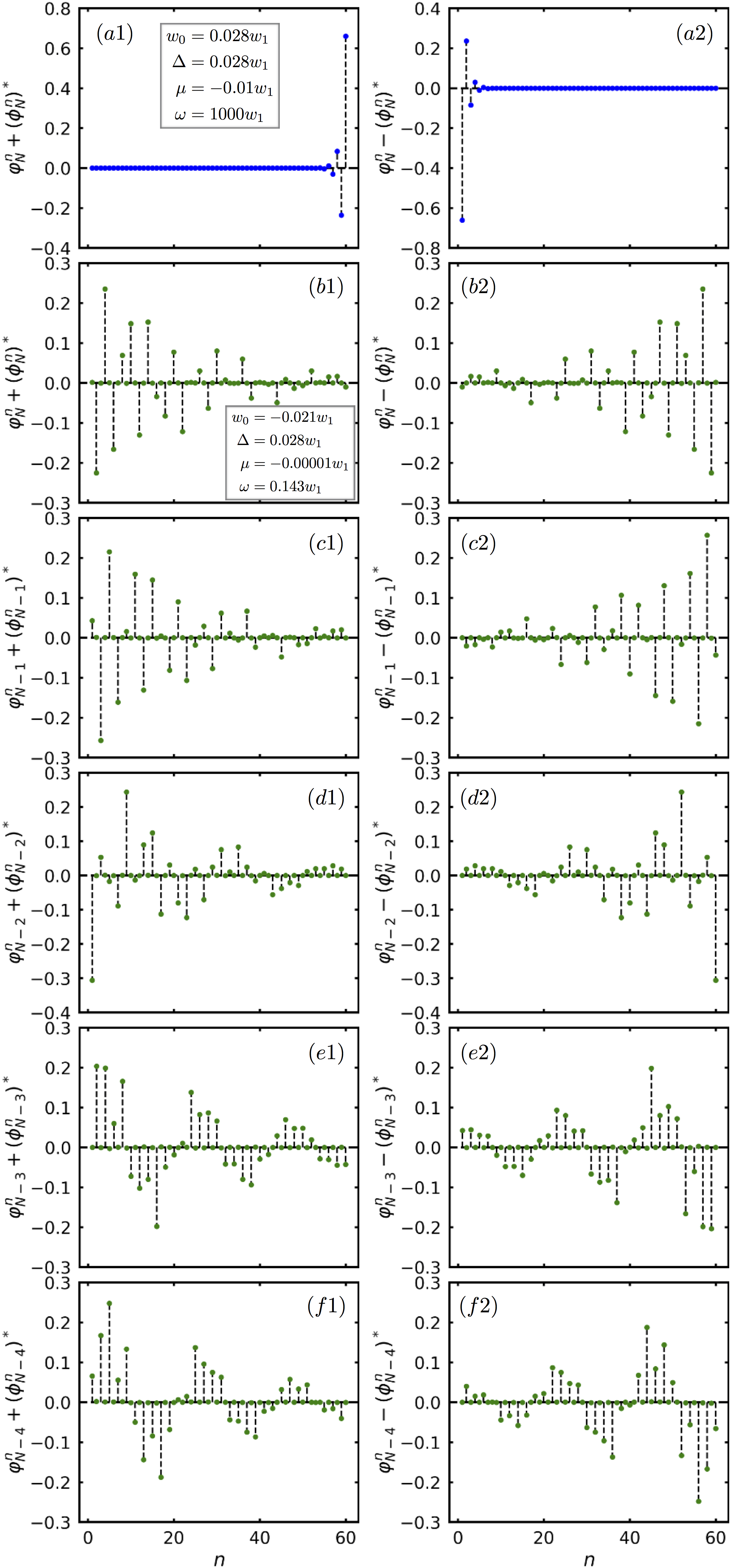} 
\caption{The spatial profile $\varphi^n_{N-i}+(\phi^n_{N-i})^*$ and $\varphi^n_{N-i}-(\phi^n_{N-i})^*$ with $i=0,1,2,3,4$ and $N=60$ of Majorana operators $\gamma_{n,\pm}$ versus the lattice index $n$. (a1) and (a2) are for a single Majorana pair corresponding to zero-energy modes i.e., $E_{60}=E_{61}=0$ in Fig.~\ref{fig:MBSspectrum}(a) while (b1)-(f2) for multiple pairs of Floquet Majorana states corresponding to five pairs of modes close to zero energy in Fig.~\ref{fig:MBSspectrum}(b).}
\label{fig:MBSwf}
\end{figure}

Figure~\ref{fig:MBSsolution} shows how coupling strengths $\mathcal{F}_t$ and $\mathcal{G}_t$ vary as $w_1/\omega$ increases. We find that there are solutions to the above-mentioned equations, i.e., $\mathcal{F}_t\mathcal{G}_t=0$ and $\mathcal{F}_t+\mathcal{G}_t\neq0$, as indicated by dots in Fig.~\ref{fig:MBSsolution}(a). The color of the dot denotes the number of Floquet Majorana pairs corresponding to the range of the {\it p}-wave pairing interaction. Figure~\ref{fig:MBSsolution}(a) shows clearly that a single pair of Majoranas i.e., $(\gamma_{1}, \gamma_{2N})$ indicated by the blue dot exists in the high-frequency limit, i.e., $w_1/\omega\rightarrow0$. This single Majorana pair is further confirmed by the zero-energy modes of the effective Hamiltonian [i.e., Eq.~(\ref{eq:rs_H_eff})], as demonstrated in Fig.~\ref{fig:MBSspectrum}(a), and also by the exponential decay of the spatial profile of Majorana operators in Figs.~\ref{fig:MBSwf}(a1) and (a2).
In addition, we find that multiple Majorana pairs are not stable. For example, nine pairs of Majoranas $(\gamma_{2l}, \gamma_{2N-(2l-1)})$ with $1\leq l\leq 9$ represented by the purple dot in Fig.~\ref{fig:MBSsolution}(a) would be destroyed by the inevitable interaction e.g., $i \gamma_{2j}\gamma_{2j+1}$ (or $i \gamma_{2j-1}\gamma_{2j+2}$) due to $\mathcal{F}_9<\mathcal{G}_1$ (or $\mathcal{F}_9<\lvert\mathcal{F}_1\rvert$) at the corresponding $w_1/\omega$.

In order to have stable Majorana pairs, an additional condition e.g., $\lvert\mathcal{G}_t\rvert>
\lvert\mathcal{G}_{t'}\rvert$ ($t'<t$) and $\lvert\mathcal{G}_t\rvert>
\lvert\mathcal{F}_{t''}\rvert$ ($t''\neq t$) when $\mathcal{F}_t=0$ or $\lvert\mathcal{F}_t\rvert>
\lvert\mathcal{F}_{t'}\rvert$ ($t'<t$) and $\lvert\mathcal{F}_t\rvert>
\lvert\mathcal{G}_{t''}\rvert$ ($t''\neq t$) when $\mathcal{G}_t=0$ should be fulfilled. 
By using parameters $w_0=-0.021\omega_1$ and $\mu=-0.00001w_1$ different from those in Fig.~\ref{fig:MBSsolution}(a), it is shown in Fig.~\ref{fig:MBSsolution}(b) that the green dot indicates $\mathcal{G}_5=\mathcal{F}_3=\mathcal{G}_1=0$ and $\mathcal{F}_5>\lvert\mathcal{G}_3\rvert> \mathcal{F}_1$, implying that there might be five pairs of Floquet Majorana states. Using the corresponding driving field, i.e., $w_1/\omega=6.989$ for the green dot, the energy spectrum of the effective Hamiltonian [i.e., Eq.~(\ref{eq:rs_H_eff})] in Fig.~\ref{fig:MBSspectrum}(b) does show multiple Floquet Majorana pairs with $E_{\nu}\sim0$.

To further support the existence of multiple pairs of Floquet Majorana states, 
we calculate the spatial profile i.e., $\varphi_{N-i}^{n}+(\phi_{N-i}^{n})^*$ and $\varphi_{N-i}^{n}-(\phi_{N-i}^{n})^*$ ($i=0,1,2,3,4$ and $n=1,2,\cdots,N$) of Majorana operators $\gamma_{n,+}=\sum_j [\varphi_j^{n}+(\phi_j^{n})^*] f_j + [\phi_j^{n} + (\varphi_j^n)^*] f_j^{\dagger}$ and $\gamma_{n,-}=-i\sum_j \{[\varphi_j^{n}-(\phi_j^{n})^*] f_j + [\phi_j^{n} - (\varphi_j^{n})^*] f_j^{\dagger}\}$~\cite{Li14PRB}. Here $\varphi_j^n$ and $\phi_j^n$ are coefficients used to define the operators $\psi_n=\sum_j \varphi_j^n f_j +\phi_j^n f_j^{\dagger}$ which together with $\psi_n^{\dagger}$ diagonalize the effective Hamiltonian in Eq.~(\ref{eq:rs_H_eff}), namely, $\tilde{H}^{\mathrm{eff}} =\sum_n E_n(\psi_n^{\dagger}\psi_n - \psi_n\psi_n^{\dagger})/2$. Figures~\ref{fig:MBSwf}(a1) and (a2) show the spatial profile  (i.e., $\varphi_{N}^{n}+(\phi_{N}^{n})^*$ and $\varphi_{N}^{n}-(\phi_{N}^{n})^*$ with $N=30$) of a single Majorana pair corresponding to the energy spectrum in Fig.~\ref{fig:MBSspectrum}(a) and the blue dot in Fig.~\ref{fig:MBSsolution}(a). 
In contrast to this well-known exponential decay starting from the edges in Fig. \ref{fig:MBSwf}(a1) and (a2), a non-exponential decay of spatial profiles of Majorana operators 
is observed in Figs.~\ref{fig:MBSwf}(b1)-(f2). The Majorana edge states for example localized at sites $2$ and $59$ decay to the bulk as shown in Fig.~\ref{fig:MBSwf}(b1) and (b2) respectively. Note that the revival behavior during the decay process in e.g., Fig.~\ref{fig:MBSwf}(d1) and (d2) is associated with a relatively large deviation from an exact zero energy in Fig.~\ref{fig:MBSspectrum}(b). These spatial profiles of Majorana operators could support the existence of multiple pairs of Floquet Majorana states as suggested already by zero-energy modes in the energy spectrum in Fig.~\ref{fig:MBSspectrum}(b) and the green dot in Fig.~\ref{fig:MBSsolution}(b). 

It is expected that many more pairs of Majoranas could also be achievable by choosing different parameters. 
We emphasize that our reported conditions for the existence of multiple Floquet Majorana edge states can  be realized by properly tuning the parameters such as the driving frequency $\omega$ and the  amplitude $w_1$ of the applied ac field. 

\begin{figure*}
\centering
  \includegraphics[width=1.8\columnwidth]{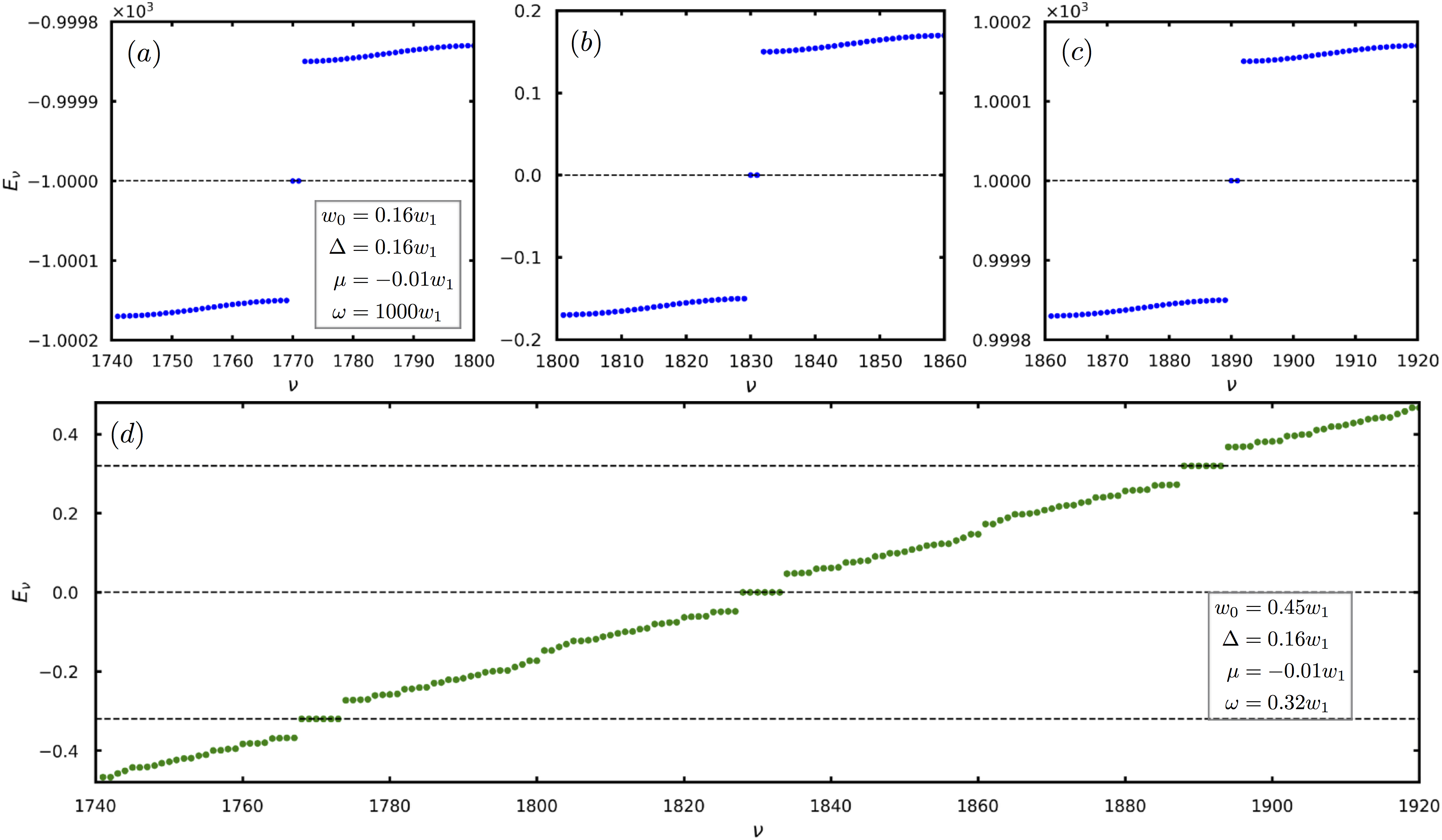} 
\caption{The quasienergy spectrum $E_{\nu}$ of the Floquet Hamiltonian [i.e., Eq.~(\ref{eq:FloquetH})] versus the index $\nu$. We take $w_1/\omega=0.001$ and $w_0=0.16w_1$ in (a), (b), (c) and $w_1/\omega=3.125$ and $w_0=0.45w_1$ in (d). Other parameters are $\Delta=0.16w_1$, $\mu=-0.01w_1$,  $N=30$, and $m=0, \pm1,\cdots, \pm30$ in Eq.~(\ref{eq:FloquetEQ}).}
\label{fig:MBSspectrumNumeric}
\end{figure*}

\subsection{Floquet spectrum}

The Magnus expansion up to a given order might give artifacts such as a dependence of the quasienergy spectrum on the driving phase or the initial time $t_0$ considered for the calculation of the one-period time evolution operator $U(t_0+T,t_0)$. This problem can be overcome by using the van Vleck expansion~\cite{EckardtAnisimovas15NJP} or the Brillouin-Wigner theory~\cite{MikamiOkaAoki16PRB}. The consideration of the van Vleck expansion up to the first order [i.e. considering only the first two terms in Eq.~({\ref{eq:effectiveH}})] gives the same results (e.g., an existence of a single Majorana pair) as those in this paper in the high frequency limit, because the commutator $[\tilde{H}_{k,-1}, \tilde{H}_{k,1}]$ equals to zero as shown in Eq. (A10).
It is nevertheless true that the driving-phase dependence in the Magnus expansion up to the first order, that does not cause any change of the spectrum within the order $\propto 1/\omega$, can contribute to finite corrections to the quasienergy spectrum~\cite{EckardtAnisimovas15NJP} after including the second-order terms ($\propto 1/\omega^2$). 
This implies that our consideration of the communicators i.e., $[\tilde{H}_{k,0}, \tilde{H}_{k,1}]$ and $[\tilde{H}_{k,0}, \tilde{H}_{k,-1}]$ in Eq. (13) could provide indications of an existence of multiple Majorana pairs. 

In order to verify our result on multiple Majorana pairs induced by a low-frequency driving, we numerically calculate the quasienergy spectrum of the time-dependent Hamiltonian by using the Floquet theory~\cite{BenitoPlatero14PRB,Shirley65PR,Sambe73PRA}. Inserting the Floquet state $\vert\psi(t)\rangle=e^{-i\epsilon t}\vert\zeta(t)\rangle$ into the Schr\"{o}dinger equation $i\partial_t\vert\psi(t)\rangle=\mathcal{H}(t)\vert\psi(t)\rangle$, we have the Floquet equation $H^{\rm F}\vert\zeta(t)\rangle=\epsilon\vert\zeta(t)\rangle$ with the Floquet Hamiltonian given by
\begin{eqnarray}
H^{\rm F}=\mathcal{H}(t)-i\partial_t. \label{eq:FloquetH}
\end{eqnarray}
In an expanded Hilbert space where $\vert\zeta(t)\rangle=\sum_m e^{im\omega t}\vert \xi_m\rangle$~\cite{Sambe73PRA}, the Floquet equation becomes 
\begin{eqnarray}
\sum_{m'} H_{m,m'}^{\rm F}\vert\xi_{m'}\rangle=\epsilon\vert\xi_m\rangle, \label{eq:FloquetEQ}
\end{eqnarray}
where  
\begin{eqnarray}
H_{m,m'}^{\rm F}&=&m\omega\delta_{m,m'}+\frac{1}{T}\int_0^T dt \mathcal{H}(t) e^{i(m'-m)\omega t}. \label{eq:FloquetHmatrix}
\end{eqnarray}
Here $\mathcal{H}(t)=\mathcal{H}_0+\mathcal{H}_{\rm D}\cos(\omega t)$ and is obtained by rewriting the time-dependent Hamiltonian [i.e., Eq.~(\ref{eq:realH})] as $H(t)=\frac{1}{2}{\bm f}^{\dagger}\mathcal{H}(t){\bm f}$ with ${\bm f}^{\dagger}=(f_1^{\dagger},\cdots,f_N^{\dagger},f_1,\cdots,f_N)$, and then the Floquet matrix elements become $H^{\rm F}_{m,m}=m\omega+\mathcal{H}_0$ and $H^{\rm F}_{m,m+1}=H^{\rm F}_{m+1,m}=\frac{1}{2}\mathcal{H}_{\rm D}$. An example of $\mathcal{H}$ with $N=4$ is provided in Appendix~\ref{sec:AppendB}. 

The Floquet quasienergy spectrum can be obtained from the numerical diagonalization of the Floquet Hamiltonian with matrix elements given by Eq.~(\ref{eq:FloquetHmatrix}). In this calculation, a small number of sites, i.e., $N=30$ (instead of $N=60$ in Figs.~\ref{fig:MBSspectrum} and \ref{fig:MBSwf}) and $m=0,\pm1,\cdots,\pm M$ with $M=30$ are considered. The Floquet Hamiltonian is then a $2N^{\prime}$-by-$2N^{\prime}$ matrix with $N^{\prime}=N(2 M+1)=1830$. The resulting energy spectrum $E_{\nu}$ is presented in Fig.~{\ref{fig:MBSspectrumNumeric}}. In the high driving frequency limit, three intervals in the  spectrum are shown in Figs.~{\ref{fig:MBSspectrumNumeric}}(a), (b), and (c) respectively. A single pair of Majorana states at the quasienergy $\epsilon=\alpha\omega$ ($\alpha=0,\pm1,\pm2,\cdots$ with $\omega=1000w_1$) appears in the spectrum. 
When a driving frequency falls short of the high frequency limit, Figure~{\ref{fig:MBSspectrumNumeric}}(d) clearly shows three pairs of Majorana edge states exactly at the quasienergy $\epsilon=\alpha\omega$ ($\alpha=0,\pm1,\pm2,\cdots$ with $\omega=w_1/3.125$) in each case. 

\begin{figure}
\centering
  \includegraphics[width=.93\columnwidth]{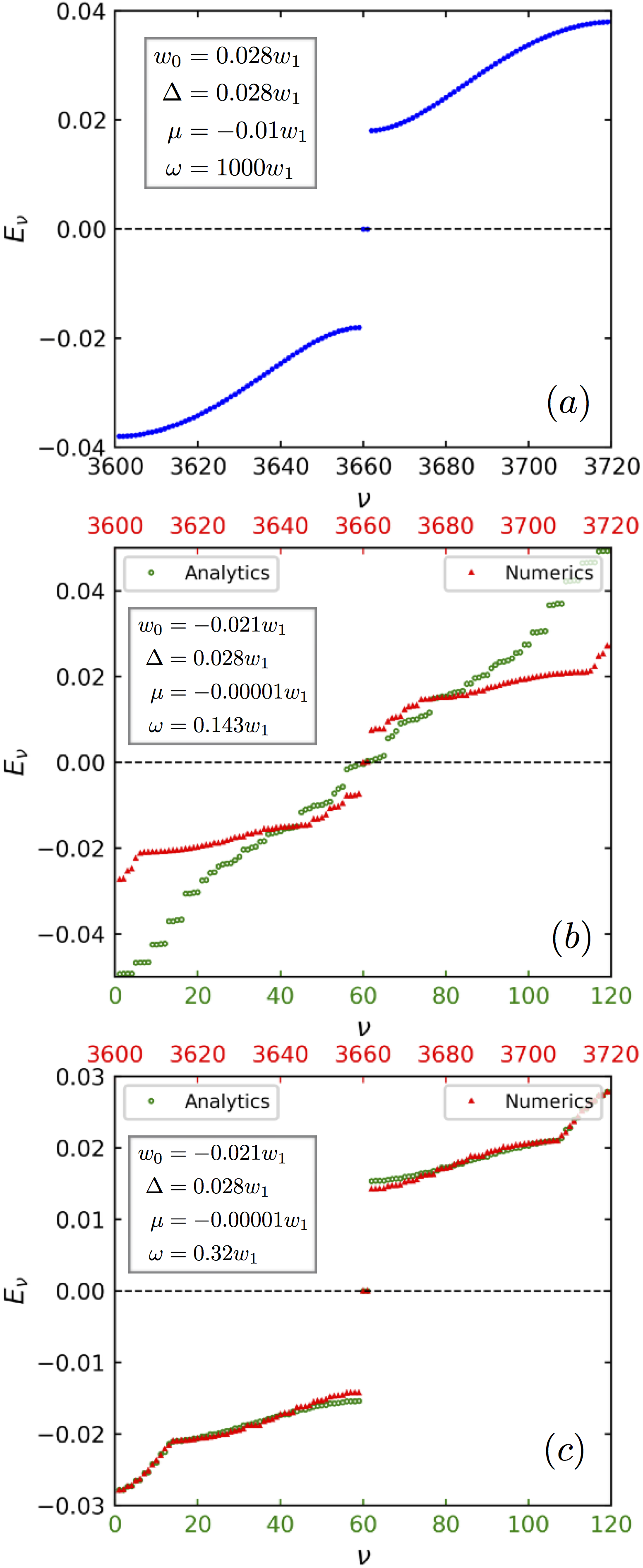} 
\caption{(a) The energy spectrum $E_{\nu}$ from numerical calculation with the same parameters as those for analytics in Fig.~\ref{fig:MBSspectrum}(a) in the high frequency limit e.g., $w_1/\omega=0.001$. (b) and (c) The comparison of energy spectra from analytics (green circles) and numerics (red upper angles) with the increase of the driving frequency from $w_1/\omega=6.989$ in (b) to $w_1/\omega=3.125$ in (c), and other parameters are same as those in Fig.~\ref{fig:MBSspectrum}(b). Note that we consider the same number of lattice sites $N=60$ as that in Fig.~\ref{fig:MBSspectrum} which is different from $N=30$ in Figs.~\ref{fig:MBSspectrumNumeric} and~\ref{fig:MBSwfNumeric}.}
\label{fig:spectrumAnaNum}
\end{figure}

Although parameters with different values from those used in Fig.~\ref{fig:MBSspectrum} are considered in Fig.~{\ref{fig:MBSspectrumNumeric}}, results in Fig.~{\ref{fig:MBSspectrumNumeric}}(b) would become identical to those in Fig.~\ref{fig:MBSspectrum}(a) when the same parameters, e.g., $w_0=\Delta=0.028w_1$ 
are used, as demonstrated in Fig.~{\ref{fig:spectrumAnaNum}}(a), indicating that our numerical calculation gives the same results as those from analytical derivations in the high frequency limit (e.g., $w_1/\omega=0.001$). 
These different parameters are, however, needed for the quasienergy spectrum to show three pairs of Majorana states in Fig.~{\ref{fig:MBSspectrumNumeric}}(d). The deviation of the conditions for the existence of multiple Majorana pairs in Fig.~{\ref{fig:MBSspectrumNumeric}}(d) from those in Fig.~\ref{fig:MBSspectrum}(b) is  not only due to neglecting terms $\propto\frac{1}{p\omega}$ with ($p=\pm2,\cdots$) in the Magnus expansion in Eq.~(\ref{eq:effectiveH}). It is also due to neglecting the second-order terms ($\propto 1/\omega^2$) that do contribute to the driving-phase-independent second-order corrections of the quasienergy spectrum~\cite{EckardtAnisimovas15NJP}. Therefore, it would be expected that this deviation decreases when increasing the driving frequency [note that numerical and analytical results in the high frequency limit are identical, as shown in Figs.~\ref{fig:MBSspectrum}(a) and~{\ref{fig:spectrumAnaNum}}(a)]. We now provide an evidence of the above explanation of the deviation by comparing the energy spectra from analytical and numerical calculations with same parameters. It is shown that the increase of the driving frequency $\omega$, e.g., from $w_1/6.989$ in Fig.~{\ref{fig:spectrumAnaNum}}(b) to $w_1/3.125$ in Fig.~{\ref{fig:spectrumAnaNum}} (c) does reduce the deviation between analytics and numerics, although these new parameters only support a single Majorana pair. Finally we want to emphasize that our analytical calculations, although not giving quantitatively precise conditions, are necessary to identify regimes where interesting effects occur and to better understand the underlining conditions  (e.g., when two terms in the effective Hamiltonian cancel at fine-tuned points).

\begin{figure}
\centering
  \includegraphics[width=.93\columnwidth]{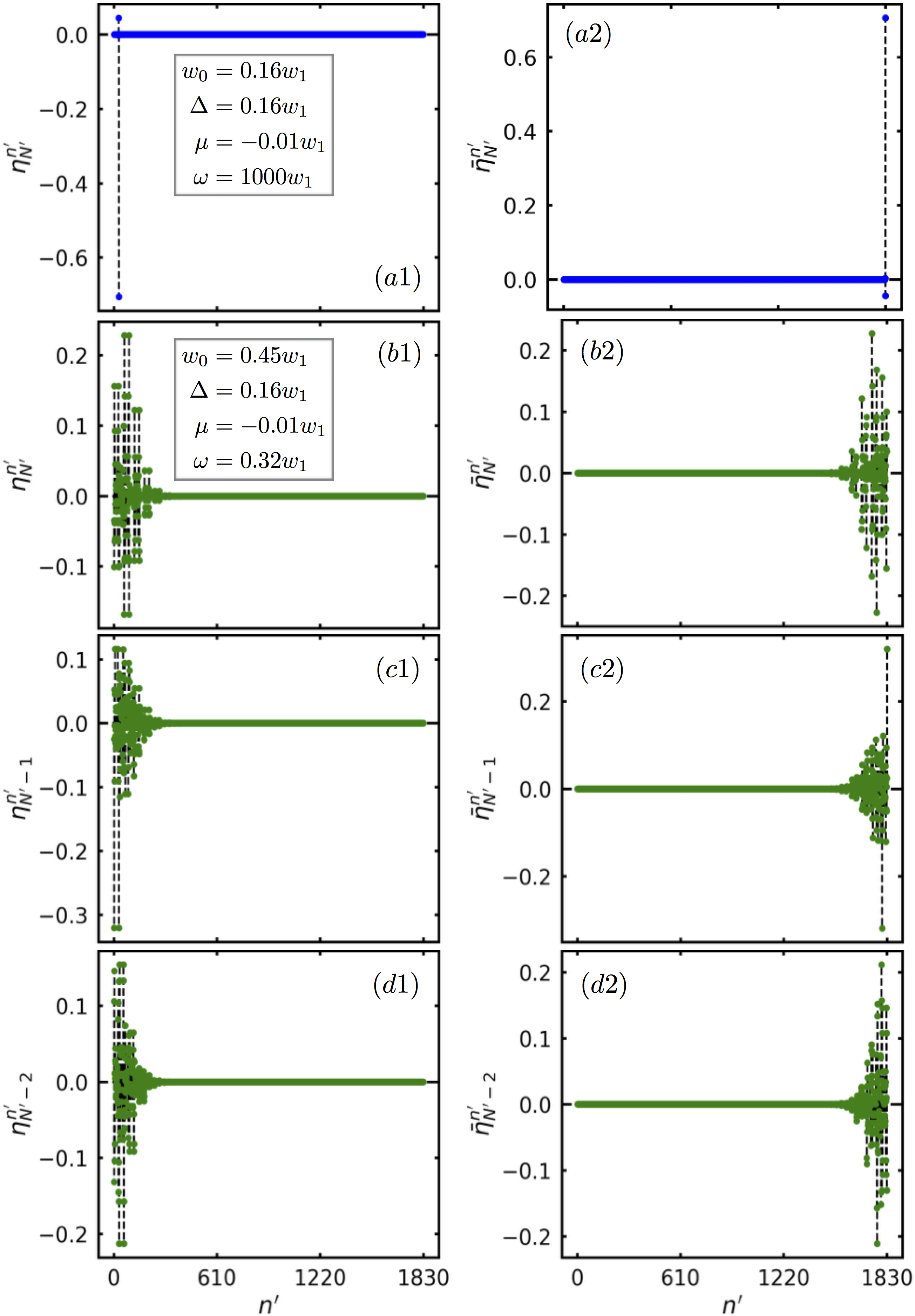} 
\caption{The spatial profile $\eta^{n^{\prime}}_{N^{\prime}-i}$ and $\bar{\eta}^{n^{\prime}}_{N^{\prime}-i}$ with $i=0,1,2$ and $N^{\prime}=1830$ of Floquet Majorana states versus the index $n^{\prime}$. (a1) and (a2) are for two zero-energy modes i.e., $E_{1830}=E_{1831}=0$ in Fig.~\ref{fig:MBSspectrumNumeric}(b) while (b1)-(d2) correspond to three pairs of zero-energy modes in Fig.~\ref{fig:MBSspectrumNumeric}(d). Other parameters as shown in the panels are same as those in Fig.~\ref{fig:MBSspectrumNumeric}.}
\label{fig:MBSwfNumeric}
\end{figure}

Similar to Fig.~{\ref{fig:MBSwf}}, Fig.~{\ref{fig:MBSwfNumeric}} shows the spatial profile (i.e., $\eta^{n^{\prime}}_{N^{\prime}-i}$ and $\bar{\eta}^{n^{\prime}}_{N^{\prime}-i}$ with $i=0,1,2$ and $N^{\prime}=1830$ obtained from eigenfunctions of the Floquet Hamiltonian) of the zero-energy modes with $E_{\nu}=0$ [see Fig.~{\ref{fig:MBSspectrumNumeric}}(d)]. Besides a single Majorana pair in the high driving frequency limit in Fig.~{\ref{fig:MBSwfNumeric}}(a1) and(a2), three pairs of well-localized Majorana edge states induced by a low-frequency driving are also observed in Fig.~{\ref{fig:MBSwfNumeric}}(b1)-(d2), verifying the results of zero-energy modes in Fig.~{\ref{fig:MBSspectrumNumeric}}(d). 
In a word, the Floquet quasienergy spectrum and the spatial profiles of the Floquet Majorana states do provide an evidence of the existence of multiple pairs of Majorana edge states when considering a driving field deviated from the high frequency limit.

\section{Discussions and conclusions \label{sec:conclusion}}

Given the known correspondence between the Kitaev chain and the one dimensional transverse Ising model~\cite{NiuChakravarty12PRB}, there is  an effective spin representation for our effective Hamiltonian obtained above.  
Specifically, we apply the inverse Jordan-Wigner transformation~\cite{Lieb61AnnPhys,DeGottardiSen13PRB} $\sigma_j^{+} = f^{\dagger}_j \prod_{m=1}^{j-1} (2f^{\dagger}_m f_m-1)$ and $\sigma_j^{-} = \prod_{m=1}^{j-1} (2f^{\dagger}_m f_m-1)f_j$ on Eq.~(\ref{eq:rs_H_eff}) and obtain an equivalent Hamiltonian in terms of spin operators, namely, 
\begin{eqnarray}
\tilde{H}^{\mathrm{eff}} &=& \sum_{j} \Big\{-\frac{\mu}{2} \sigma_j^{x} 
- \frac{w_{0}}{4} (\sigma_j^y \sigma_{j+1}^y +\sigma_j^z \sigma_{j+1}^z) \notag \\ 
&& 
+ \frac{\mu  \Delta}{\omega} 
(\sigma_j^z \sigma_{j+1}^z -\sigma_j^y \sigma_{j+1}^y ) \notag \\
&&
+\sum_{m=0}^{\infty} \sum_{r=1,3,\cdots}^{2m+3} 
\frac{\mathcal{C}_1\mathcal{D}_1}{2} 
( \sigma_j^y M_{j,r}^{x} \sigma_{j+r}^y 
+\sigma_j^z M_{j,r}^{x} \sigma_{j+r}^z ) \notag \\ 
&& 
-\sum_{m=0}^{\infty} \sum_{r^{\prime}=1,3,\cdots}^{2m+1} 
\frac{\mathcal{C}_2\mathcal{D}_2}{2} 
(\sigma_j^z M_{j,r^{\prime}}^{x} \sigma_{j+r^{\prime}}^z - \sigma_j^y M_{j,r^{\prime}}^{x} \sigma_{j+r^{\prime}}^y) \notag \\
&&
- \sum_{m=0}^{\infty} \sum_{r=1,3,\cdots}^{2m+3} \frac{\mathcal{C}_3 \mathcal{D}_3}{2} 
(\sigma_j^z M_{j,r}^{x} \sigma_{j+r}^z - \sigma_j^y M_{j,r}^{x} \sigma_{j+r}^y)
\Big\}, \notag \\
\label{eq:spinH_tunneling}
\end{eqnarray}
where
\begin{eqnarray}
M_{j,\chi}^{x}&=&\prod_{m=j+1}^{j+\chi-1}\sigma_{m}^{x},  \text{  } \chi=r, r^{\prime}.
\end{eqnarray} 
It shows long-range many-body spin interactions with strengths $\mathcal{C}_1\mathcal{D}_1/2$, $\mathcal{C}_2\mathcal{D}_2/2$, and $\mathcal{C}_3\mathcal{D}_3/2$ that correspond to long-range tunneling and pairing interactions in Eq.~(\ref{eq:rs_H_eff}). Note that the spin interactions involving $\mathcal{C}_1$ and $\mathcal{C}_3$, and the nearest-neighbor spin interaction with the strength $\mu\Delta/\omega$ are leading corrections to the high-frequency limit. 

The multiple pairs of Floquet Majorana modes have been demonstrated previously in the entire region of the phase diagram in Ref.~\onlinecite{Thakurathi13PRB} where the periodic $\delta$-function kicks in the chemical potential and hopping terms were considered. The multiple Majorana pairs presented in our paper are, however, based on a different driving protocol, namely harmonically driven tunneling interactions. We have verified (but not included in this paper) that multiple Majorana pairs together with long-range interactions do not appear when considering the harmonic driving of the chemical potential~\cite{BenitoPlatero14PRB}, which is consistent with numerical results of two end states in Ref.~\onlinecite{Thakurathi13PRB}. Our paper shows the existence of multiple pairs of Majorana edge states for certain parameters and an extension of these isolated points to entire regions of phase diagram might be possible. 

For a Kitaev chain with or without high frequency driving, there would be two edge states. The overlap of their wavefunctions results in a Majorana splitting $\propto e^{-L/\xi}$ where $L$ is the length of the chain and $\xi$ is the correlation length. When considering a driving falling short of the high frequency limit, we have shown that several edge states at each end of the chain appear. To make the Majorana splitting negligible compared with all relevant energy scales, two nearest-neighboring edge states should be sufficiently well separated. The detail investigation of Majorana mode splitting of our effective long-range interacting model will be performed in our future work. 

We believe that it is possible to observe experimentally the effects of the driving frequency away from the high frequency limit on the Floquet engineering of long-range {\it p}-wave superconductivity and Floquet Majorana edge states deduced in this paper. This is because the tunneling of carriers can in general be tuned via the gate voltages in quantum-transport experiments. Possible physical realizations include a ferromagnetic atomic chain on a superconductor~\cite{NadjBernevigYazdani14science}, a one-dimensional wire with Rashba spin-orbit interaction~\cite{LutchynSauSarma10PRL}, and a semiconductor quantum dot coupled to superconducting grains~\cite{SauSarma12NatComm}.

We have studied the effects of the leading correction in terms of the inverse driving frequency on the Floquet engineering of long-range {\it p}-wave superconductivity in a Kitaev chain. We find that the leading corrections can generate new long-range {\it p}-wave pairing interactions, which could appreciably correct the ones present at the high-frequency limit when the interaction range increases. We also find that long-range tunneling interactions and nearest-neighbor {\it p}-wave pairings can be generated when applying a broad range of driving frequencies. In addition, we show the leading corrections can give detectable multiple pairs of Floquet Majorana edge states that is destroyed at the high driving frequency limit. 

\acknowledgements
ZZL thanks Yong-Chang Zhang and Sheng-Wen Li for very helpful discussions. We thank the anonymous referees for constructive criticisms that helped further improve this paper. We acknowledge support from the National Natural Science Foundation of China (Grant No.~11404019) and HK PolyU (Grant No.~G-YBHY).

\appendix

\section{Derivation of Eq.~(\ref{eq:bdg_tunnel_Bessel}) \label{sec:AppendA}}
The Fourier components given in Eq.~(\ref{eq:FourierComp-tunn}) for $p=0,\pm 1$ become 
\begin{eqnarray}
\tilde{H}_{k,0}&=&\mathcal{A}\sigma_k^z-i\mathcal{B}\mathcal{J}_{0}\sigma_k^{+}+i\mathcal{B}\mathcal{J}_{0}\sigma_k^{-},  \\
\tilde{H}_{k,1}&=&-i\mathcal{B}\mathcal{J}_{-1}\sigma_k^{+}+i\mathcal{B}\mathcal{J}_{1}\sigma_k^{-},  \\
\tilde{H}_{k,-1}&=&-i\mathcal{B}\mathcal{J}_{1}\sigma_k^{+}+i\mathcal{B}\mathcal{J}_{-1}\sigma_k^{-}, 
\end{eqnarray}
where
\begin{eqnarray}
\mathcal{A} &=& \mu- w_{0}\cos k, \\
\mathcal{B} &=& \Delta \sin k, \\
\mathcal{J}_0 &\equiv& \mathcal{J}_0\left( \frac{w_{1}}{\omega} \cos k \right), \\
\mathcal{J}_{\pm1} &\equiv& \mathcal{J}_{\pm1} \left( \frac{w_{1}}{\omega} \cos k \right).
\end{eqnarray}
Their commutators are obtained as
\begin{eqnarray}
\left[\tilde{H}_{k,0},\tilde{H}_{k,1}\right] &=& -2\mathcal{A}\mathcal{B} \mathcal{J}_{1} \sigma_k^{y} 
 +2\mathcal{B}^2 \mathcal{J}_0 \mathcal{J}_{1} \sigma_k^z, \\
\left[\tilde{H}_{k,0},\tilde{H}_{k,-1}\right] &=& 2\mathcal{A}\mathcal{B} \mathcal{J}_{1} \sigma_k^{y} 
 -2\mathcal{B}^2 \mathcal{J}_0 \mathcal{J}_{1} \sigma_k^z, \\
\left[\tilde{H}_{k,1},\tilde{H}_{k,-1}\right] &=& 0.
\end{eqnarray}
Here, we have used $[\sigma_k^z, \sigma_k^+]=2\sigma_k^+$, $[\sigma_k^z, \sigma_k^-]=-2\sigma_k^-$, $[\sigma_k^+, \sigma_k^-]=\sigma_k^z$, and $\mathcal{J}_{-1}=-\mathcal{J}_{1}$. Inserting these commutators into Eq.~(\ref{eq:effectiveH}), we have
\begin{eqnarray}
\tilde{H}_k^{\mathrm{eff}} &=& \tilde{H}_{k,0} +\frac{1}{\omega} \{[\tilde{H}_{k,0},\tilde{H}_{k,1}] -[\tilde{H}_{k,0},\tilde{H}_{k,-1}]  \notag \\
&& +[\tilde{H}_{k,1},\tilde{H}_{k,-1}] \} \notag \\
&=& \mathcal{A}\sigma_k^z +i \mathcal{B} \left[\mathcal{J}_{0} 
-\frac{4 }{\omega} \mathcal{A} \mathcal{J}_{1} \right] ( \sigma_k^{-}-\sigma_k^{+} ) \notag \\
&& + \frac{4 }{\omega} \mathcal{B}^2 \mathcal{J}_0\mathcal{J}_{1} \sigma_k^z.
\end{eqnarray}
Further inserting the expressions of $\mathcal{A}$ and $\mathcal{B}$, the effective Hamiltonian in Eq.~(\ref{eq:bdg_tunnel_Bessel}) is then obtained. Using $\tilde{H}^{\mathrm{eff}}=\sum_{k>0}\Psi_k^{\dagger} \tilde{H}_k^{\mathrm{eff}}\Psi_k$, the Hamiltonian in the reciprocal space becomes
\begin{eqnarray}
\tilde{H}^{\mathrm{eff}} 
&=& \sum_{k>0}\Big\{ \Big(\mu - w_{0} \cos k +\frac{4\mathcal{J}_0 \mathcal{J}_{1}}{\omega} \Delta^2\sin^2 k \Big) \notag \\
 && \times \Big(f_k^{\dagger}f_k-f_{-k}f_{-k}^{\dagger}\Big) 
-i \Big[ \Big(\mathcal{J}_{0} -\frac{4\mu}{\omega} \Big)\Delta \sin k \notag \\
&& +\frac{2w_{0} \mathcal{J}_{1} }{\omega} \Delta \sin(2k)  \Big] 
\Big(f_{k}^{\dagger} f_{-k}^{\dagger} -f_{-k}f_{k} \Big) \Big\} . 
\label{eq:paritingV}
\end{eqnarray}

\section{An example of $\mathcal{H}(t)$ in Eq.~(\ref{eq:FloquetHmatrix}) \label{sec:AppendB}}

In this appendix, we provide an example of $\mathcal{H}(t)=\mathcal{H}_0+\mathcal{H}_{\rm D}\cos(\omega t)$ that is used to construct the Floquet Hamiltonian matrix in Eq.~(\ref{eq:FloquetHmatrix}). Consider a chain with four sites i.e., $N=4$, we have
\begin{eqnarray}
\mathcal{H}_0 &=&
\left( \begin{array}{cccccccc} 
\mu & -\frac{w_0}{2} & 0 & 0 & 0 & -\frac{\Delta}{2} & 0 & 0  \\
-\frac{w_0}{2} &  \mu & -\frac{w_0}{2} & 0 & \frac{\Delta}{2} & 0 & -\frac{\Delta}{2} & 0   \\
0 & -\frac{w_0}{2} &  \mu & -\frac{w_0}{2} & 0 & \frac{\Delta}{2} & 0 & -\frac{\Delta}{2}   \\
0 & 0 & -\frac{w_0}{2} & \mu & 0 & 0 & \frac{\Delta}{2} & 0   \\
0 & \frac{\Delta}{2} & 0 & 0 & -\mu & \frac{w_0}{2}  & 0 & 0   \\
-\frac{\Delta}{2} & 0 & \frac{\Delta}{2} & 0 & \frac{w_0}{2} & -\mu & \frac{w_0}{2} & 0   \\
0 & -\frac{\Delta}{2} & 0 & \frac{\Delta}{2} & 0 & \frac{w_0}{2} & -\mu & \frac{w_0}{2}   \\
0 & 0 & -\frac{\Delta}{2} & 0 & 0 & 0 & \frac{w_0}{2} & -\mu   \\
\end{array} \right) \notag \\
\end{eqnarray}
and
\begin{eqnarray}
\mathcal{H}_{\rm D} &=&
\left( \begin{array}{ccccccccc} 
0 & -\frac{w_1}{4} & 0 & 0 & 0 & 0 & 0 & 0  \\
 -\frac{w_1}{4}& 0 &-\frac{w_1}{4}  & 0 & 0 & 0 & 0 & 0  \\
0 & -\frac{w_1}{4} & 0 & -\frac{w_1}{4} & 0 & 0 & 0 & 0  \\
0 & 0 & -\frac{w_1}{4} & 0 & 0 & 0 & 0 & 0  \\
0 & 0 & 0 & 0 & 0 & \frac{w_1}{4} & 0 & 0  \\
0 & 0 & 0 & 0 & \frac{w_1}{4} & 0 & \frac{w_1}{4} & 0  \\
0 & 0 & 0 & 0 &0  & \frac{w_1}{4} & 0 & \frac{w_1}{4}  \\
0 & 0 &0  & 0 & 0 & 0 & \frac{w_1}{4} & 0  \\
\end{array} \right). \notag\\
\end{eqnarray}
The quasienergy spectrum is then obtained from the diagonalization of the Floquet Hamiltonian with matrix elements given by Eq.~(\ref{eq:FloquetHmatrix}).

\end{document}